\documentclass[sigconf,nonacm]{acmart}
\renewcommand\footnotetextcopyrightpermission[1]{}
\newcommand{\laxman}[1]{{\color{brown}{\bf Laxman:} #1}}
\newcommand{\kuba}[1]{{\color{violet}{\bf Kuba:} #1}}
\newcommand{\guy}[1]{{\color{olive}{\bf Guy:} #1}}
\newcommand{\lars}[1]{{\color{blue}{\bf Lars:} #1}}
\newcommand{\tobias}[1]{{\color{purple}{\bf Tobias:} #1}}
\newcommand{\richard}[1]{{\color{magenta}{\bf Richard:} #1}}
\definecolor{danielgreen}{RGB}{20, 125, 45}
\newcommand{\daniel}[1]{{\color{danielgreen}{\bf Daniel:} #1}}

\newcommand{\pipnn}{PiPNN\xspace}

\newcommand{\anns}{ANNS\xspace}

\newcommand{\mX}{\mathcal{X}}
\DeclareMathOperator*{\argmin}{arg\,min}

\newcommand{\eg}{e.g.,\xspace}
\newcommand{\myparagraph}[1]{\smallskip\noindent\textbf{#1.}\xspace}

\newtheorem{definition}{Definition}[section]

\usepackage{xspace}
\usepackage[linesnumbered,ruled,vlined]{algorithm2e}
\SetAlCapFnt{\small}
\SetAlCapNameFnt{\small}
\usepackage{enumitem}
\usepackage{graphicx}
\usepackage{pgf}
\usepackage{tikz}

\usepackage{multirow}
\usepackage{colortbl}
\usepackage[normalem]{ulem}

\AtBeginDocument{%
  }

\makeatletter
\def\@authorfont{\Large}
\def\@affiliationfont{\normalsize}
\makeatother

\begin{document}
\title{GINGER: Spicing Up Graph-Based Vector Search}


\author{Daniel Salwasser}
\affiliation{%
  \institution{Google Research \& KIT}
  \city{Karlsruhe}
  \country{Germany}
}
\email{dsalwasser@google.com}

\author{Lars Gottesb\"uren}
\affiliation{%
  \institution{Google Research}
  \city{Zürich}
  \country{Switzerland}
}
\email{gottesbueren@google.com}

\author{Laxman Dhulipala}
\affiliation{%
  \institution{Google Research \& UMD}
  \city{College Park}
  \country{USA}
}
\email{laxmand@google.com}

\author{Jakub {\L}{\k a}cki}
\affiliation{%
  \institution{Google Research}
  \city{New York City}
  \country{USA}
}
\email{jlacki@google.com}

\author{Guy Blelloch}
\affiliation{%
  \institution{Google Research \& CMU}
  \city{Pittsburgh}
  \country{USA}
}
\email{gblelloch@google.com}

\author{Rajesh Jayaram}
\affiliation{%
  \institution{Google Research}
  \city{New York City}
  \country{USA}
}
\email{rkjayaram@google.com}

\author{Tobias Rubel}
\affiliation{%
  \institution{Google Research \& UMD}
  \city{College Park}
  \country{USA}
}
\email{rubelato@google.com}

\author{Kshiteej Sheth}
\affiliation{%
  \institution{Google Research \& EPFL}
  \city{Zürich}
  \country{Switzerland}
}
\email{kshiteejsheth@google.com}

\author{Amir Zandieh}
\affiliation{%
  \institution{Google Research}
  \city{Zürich}
  \country{Switzerland}
}
\email{zandieh@google.com}

\author{Vahab Mirrokni}
\affiliation{%
  \institution{Google Research}
  \city{New York City}
  \country{USA}
}
\email{mirrokni@google.com}

\renewcommand{\shortauthors}{Salwasser et al.}

\begin{abstract}
We present \textsc{Ginger}, a state-of-the-art open-source C++ library for graph-based approximate nearest neighbor search (ANNS).
We developed \textsc{Ginger} at Google to push the speed and memory efficiency of graph search on large and high-dimensional embedding datasets.
\textsc{Ginger} surpasses existing ANNS libraries with regards to query throughput and indexing time while matching the memory footprint of the most space-efficient competitors.

\textsc{Ginger} builds upon recent advances in proximity graph construction (PiPNN) and vector quantization (TurboQuant). We introduce Nested TurboQuant (NTQ), a two-level quantizer which supports low precision graph navigation and high precision reranking in a single code. This representation achieves near floating point precision with $2.6\times$ less memory. We also compress the size of our proximity graphs by up to $2.5\times$ without impacting query throughput.
We combine graph reordering with a new lossless compression codec called AdaptiveGap, which builds upon StreamVByte.
At $95\%$ recall, \textsc{Ginger} speeds up query throughput by an average of $1.8\times$ over each dataset's fastest competitor and by at least $2.3\times$ over any single library.
This lead grows to $2.9\times$ over compact libraries on high-dimensional embeddings.

\end{abstract}

\maketitle

\providecommand{\sys}{\textsc{Ginger}}

\section{Introduction}
\label{sec:intro}

The rapid progress in deep learning has placed Approximate Nearest Neighbor Search (ANNS) at the center of information retrieval and data analysis~\cite{LiLiSuhara:2020,ZeakisPapadakisSkoutas:2023,LewisPerezPiktus:2020,gao2023retrieval,mitra2018introduction,roy2022systematic,XiongXLTLBAO21}.
Today's workloads are marked by a sharp increase in both the number of indexed entities and the dimensionality of the embeddings used to represent them.

Proximity graphs have been shown to perform well in very high dimensions, and are thus the dominant technique for ANNS indexing whenever high recall is paramount.
A proximity graph represents each entity as a vertex and connects it to nearby and further directionally diverse neighbors.
Queries route through the graph via greedy \emph{beam search}, which iteratively expands the closest unexplored vertices until converging to the query neighborhood.
Prominent proximity-graph algorithms include HNSW~\cite{malkov2018efficient}, Vamana (DiskANN)~\cite{jayaram2019diskann}, NSG~\cite{nsg}, and PiPNN~\cite{rubel2026pipnn,rubel2026turbocharging}.
Compared to cluster-based inverted-file (IVF) indexes such as Faiss~\cite{douze2024faiss} and ScaNN~\cite{avq_2020,soar_2023}, proximity graphs achieve high recall with many fewer distance evaluations on modern embeddings.

Still, graph indexes face three bottlenecks in practice: (1) memory bandwidth at search time, (2) memory footprint to host the index, and (3) indexing time.
Our prior work PiPNN~\cite{rubel2026pipnn, rubel2026turbocharging} effectively resolved the issue of index construction, building high-quality billion-scale indexes in 10s of minutes instead of hours. 
This paper addresses the other two bottlenecks: search throughput and memory footprint. 
Moreover, we introduce the first production-grade implementation of PiPNN.

\begin{figure}[t!]
  \centering
  \includegraphics[width=\columnwidth]{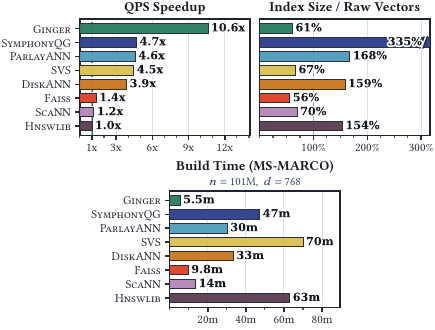}
  \caption{Query throughput relative to Hnswlib at 95\% $10@10$ recall (top left, higher is better) and memory footprint (graph and compressed vectors) relative to raw vectors (top right, lower is better) as the geometric mean across 14 datasets, alongside index construction time on \textsc{MS-MARCO} (bottom, lower is better). SymphonyQG and ScaNN fail on some datasets, which are excluded in their aggregate.
  }
  \label{fig:teaser}
\end{figure}

To this end we present and open-source \sys{}\footnote{\url{https://github.com/google/ginger}}, an in-memory C++ library for graph-based ANNS.
It was developed internally at Google to solve a variety of large-scale retrieval problems with a primary focus on efficiency.
We developed \sys{} from the ground up as a standalone C++ library with portable SIMD kernels written in Highway~\cite{highway}.
\sys{} also serves as the official open-source implementation of PiPNN II~\cite{rubel2026turbocharging}, thus combining SOTA query speed, memory efficiency and indexing time in one package.

Many industrial-grade vector search libraries exist, including Faiss~\cite{douze2024faiss}, ScaNN~\cite{avq_2020,soar_2023}, DiskANN~\cite{jayaram2019diskann}, SVS~\cite{aguerrebere2023similarity}, and Hnswlib~\cite{hnswlib}.
Given this high bar, it might be expected that there exists little room for improvement over these widely adopted and well-optimized systems.
Surprisingly, we find that a combination of algorithmic improvements in vector quantization as well as careful performance engineering yields significant speedups over the prior state of the art.
Across all 14 benchmarks in our evaluation, \sys{} is on average $1.8\times$ faster than the fastest competitor at $95\%$ recall (up to $2.7\times$).
The gap is largest on high-dimensional datasets ($d \ge 768$), where \sys{} is $2.0\times$ faster on average.
Our results come from both designing every step of our beam search implementation around reducing memory traffic as well as reducing the index footprint by way of lossless graph compression.
We make the following contributions.

\begin{itemize}[leftmargin=*]
\item \textbf{Nested TurboQuant (Section~\ref{sec:quant}).}
We introduce Nested TurboQuant (NTQ), an extension of TurboQuant in which a single code per vector serves both graph navigation and reranking.
The beam search operates on the leading 4 bits per coordinate, which form a valid TurboQuant code~\cite{zandieh2025turboquant,vargaftik2022eden}.
The remaining $k$ bits refine each 4-bit interval with a nested Lloyd--Max codebook, using the full code to rerank the search results.
With 12 bits per coordinate, NTQ nearly matches the accuracy of 32-bit floats while using $2.6\times$ less memory.

\item \textbf{Fast graph search (Sections~\ref{sec:quant} and~\ref{sec:search}).}
We contribute an extremely fast SIMD kernel for one-to-many scoring of the 4-bit codes.
On high-dimensional embeddings, we offer two ways to further speed up the search.
We either project vectors to a lower dimension with PCA, or apply a tuning-free prefilter called TurboSlice~\cite{daliri2026turboslice} to discard unpromising candidates early.

\item \textbf{Graph compression (Section~\ref{sec:compress}).}
On low-dimensional datasets, the graph adjacency lists of prior libraries consume as much memory as the quantized vectors themselves.
We introduce \emph{AdaptiveGap}, a lossless codec that packs small gaps into fewer than eight bits while retaining the single-shuffle SIMD decoding of StreamVByte.
On the same vertex ordering, AdaptiveGap uses on average $5.4\%$ less graph memory (up to $7.6\%$) than StreamVByte.
We combine this codec with graph partitioning to assign nearby IDs to neighboring vertices.
Together, these techniques shrink the graph by an average of $2.1\times$ (up to $2.5\times$) relative to an uncompressed CSR layout without slowing down queries.

\item \textbf{Extensive evaluation (Section~\ref{sec:eval}).}
We compare \sys{} with seven state-of-the-art graph and IVF libraries on 14 public benchmarks with up to one billion vectors and 4096 dimensions.
\sys{} achieves the highest query throughput at $90\%$ and $95\%$ recall on 13 of the 14 benchmarks.
At $95\%$ recall, \sys{} outperforms the fastest competitor by $2.0\times$ on average (up to $2.7\times$) on high-dimensional data and by $1.45\times$ (up to $2.1\times$) on low-dimensional data.
At $90\%$ recall, the corresponding speedups are $1.8\times$ (up to $2.3\times$) and $1.55\times$ (up to $2.0\times$).
The complete \sys{} index for the billion-point \textsc{Deep} dataset occupies approximately $288$~GB of memory with 12-bit NTQ (Figure~\ref{fig:memory_breakdown}).
\end{itemize}

\sys{}'s modular design and portable SIMD kernels make it easy to integrate state-of-the-art ANNS into new search and data management systems.
Our results demonstrate that we can \emph{simultaneously} deliver state-of-the-art memory footprint, search performance, and indexing time through a mix of algorithmic improvements and performance engineering.

\section{Preliminaries}
\label{sec:prelim}

Let $\mX \subseteq \mathbb{R}^d$ be a collection of $n$ vectors and let $\mathrm{dist}(\cdot,\cdot)$ be a dissimilarity on $\mathbb{R}^d$, such as Euclidean distance or negative inner product.

\begin{definition}[$k$-nearest-neighbor search]
Given a query $q \in \mathbb{R}^d$, the $k$-nearest-neighbors of $q$ in $\mX$ is a set $\mathcal{K} \subseteq \mX$ with $|\mathcal{K}| = k$ such that $\mathrm{dist}(q,p) \le \mathrm{dist}(q,p')$ for all $p \in \mathcal{K}$ and $p' \in \mX \setminus \mathcal{K}$.
\end{definition}

\myparagraph{Approximate Nearest Neighbors (ANNS)}
Given a query $q$, an approximate nearest neighbors (ANNS) index returns a set $\mathcal{K}'$ that approximates $\mathcal{K}$. The quality is measured by the \emph{recall}, $|\mathcal{K} \cap \mathcal{K}'| / k$, averaged over a query set. We write $k@k'$ recall when $|\mathcal{K}'| = k' \ge k$, and use $10@10$ recall unless stated otherwise. An index is judged by its throughput (queries per second, QPS) at a given recall, by its memory footprint, and by the time taken to construct it.

\subsection{\anns Indexes}

\myparagraph{Proximity Graphs and Search}
A graph-based ANNS index organizes vectors into a proximity graph, with a vertex for each vector and directed edges connecting vectors. Neighbors are selected heuristically during graph construction based on closeness and directional diversity.
Queries explore this graph using beam search (Algorithm~\ref{alg:beam_search}).
Starting from an entry vertex $s$, the search maintains a beam $B$ of at most $L$ \emph{candidates} and a set $V$ of \emph{expanded} vertices.
It repeatedly expands the closest vertex in $B \setminus V$, adds its neighbors to the beam, and retains the $L$ closest candidates, stopping when every vertex in the beam has been expanded.
Implementations of graph search differ widely in their candidate and visited-set data structures, stopping rules, filtering, batching, and prefetching.
They may also navigate using quantized vectors and rerank a larger candidate set using full-precision vectors or more accurate codes.
These choices affect memory traffic, query throughput, and recall.

\begin{algorithm}
    \caption{BeamSearch}
    \label{alg:beam_search}
    \SetKwData{beam}{$B$}
    \SetKwData{beamwidth}{$L$}
    \SetKwData{visited}{$V$}
    \SetKwData{graph}{$G$}
    \SetKwData{source}{$s$}
    \SetKwData{target}{$q$}
    \SetKwData{points}{$\mX$}
    \SetKwData{curr}{$p$}
    \SetKwFunction{FNeighbors}{Neighbors}
    \KwIn{$\graph = (\points, E)$ a proximity graph over $\points$, $\source \in \points$ an entry point, $\target$ a query point, and $\beamwidth \in \mathbb{N}$ a beam width}
    \KwOut{$\beam \subseteq \points$ containing the $\beamwidth$ closest points to $\target$ encountered during the search}
    $\beam \gets \{\source\}$\;
    $\visited \gets \varnothing$\;
    \While{$\beam \setminus \visited \neq \varnothing$}{
        $\curr \gets \argmin_{u \in \beam \setminus \visited} \mathrm{dist}(u,\target)$\;
        $\beam \gets \beam \cup \FNeighbors{\curr}$\;
        $\visited \gets \visited \cup \{\curr\}$\;
        \If{$|\beam| > \beamwidth$}{
            $\beam \gets$ the $\beamwidth$ points in $\beam$ with smallest distance to $\target$\;
        }
    }
    \Return \beam\;
\end{algorithm}

\myparagraph{Graph Construction}
The \textbf{Vamana}~\cite{jayaram2019diskann} index constructs a proximity graph iteratively via beam-search traversals on the partial index and an $\alpha$-Relative Neighborhood Graph ($\alpha$-RNG) pruning heuristic (\texttt{RobustPrune}). This rule enforces directional diversity and injects long-range edges, achieving shorter search paths than hierarchical structures like HNSW~\cite{malkov2018efficient}.
Many other proximity-graph construction algorithms have been proposed~\cite{dong2011efficient,malkov2014approximate,fu2016efanna,harwood2016fanng,iwasaki2018optimization,nsg,munoz2019hierarchical,fu2021high,zhao2023towards,yang2024revisiting,yu2025pecann,voruganti2025mirage}.
In practice, Vamana and HNSW are the most widely used.
However, search-driven graph construction suffers from a fundamental ``search bottleneck''~\cite{rubel2026pipnn} requiring millions of random-access memory lookups and thus incurring long build times.

\textbf{PiPNN} (Pick-in-Partitions Nearest Neighbors)~\cite{rubel2026pipnn} speeds up index construction by replacing searches during construction with a partitioning-based build paradigm. Instead of global beam searches, PiPNN recursively decomposes the dataset into overlapping clusters via randomized ball carving and generates candidate edges using cache-friendly many-to-many top-$k$ kernels. This achieves an order-of-magnitude faster index construction across CPUs and accelerators while maintaining equivalent graph quality.
In \sys{}, we offer both \textsc{PiPNN} and \textsc{Vamana} index builds to give users flexibility and choice.

\myparagraph{Inverted Indexes}
Inverted-file (IVF) indexes partition vectors into clusters and store each cluster in an inverted list.
A query probes a subset of the clusters and scores the vectors in their lists, often using quantized representations followed by higher-precision reranking.

\myparagraph{Quantization}
Quantization replaces floating-point vectors with compact codes representing regions of the vector space, reducing both index memory and the bytes fetched during distance evaluation~\cite{jegou2011product,aguerrebere2023similarity}.
The codes must support fast scoring while preserving enough accuracy to identify the nearest neighbors.
\emph{Scalar quantization} encodes each coordinate independently using $b$ bits to select one of $2^b$ representative values.
\emph{Product quantization} partitions a vector into subvectors and encodes each with an index into a learned codebook; distances are estimated by combining per-subvector table lookups~\cite{jegou2011product,ge2013optimized}.
Rotation-based scalar quantizers first apply a shared random orthogonal transformation, then encode the transformed coordinates.
Examples include DRIVE/EDEN~\cite{vargaftik2021drive,vargaftik2022eden,benbasat2026eden}, RaBitQ~\cite{gao2024rabitq,gao2025rabitq,gao2025extendedrabitq}, and TurboQuant~\cite{zandieh2025turboquant}.
The representation and scoring kernel must also suit the search access pattern.
Graph traversal scores scattered candidates, whereas IVF search scans lists of vectors.
Quantization has a rich literature, and we refer readers to the excellent surveys~\cite{gray1998quantization,matsui2018survey} for background on its foundations and applications to nearest-neighbor search.
Section~\ref{sec:quant} presents our nested quantizer and SIMD scoring kernels for graph navigation and reranking.

\subsection{Prior Libraries}\label{sec:prior-libraries}

We focus on a representative selection of widely used libraries and high-performance ANNS methods.
Our selection emphasizes methods that are widely adopted and/or very high performance while covering graph-based and inverted-file search.

\myparagraph{DiskANN}
DiskANN~\cite{jayaram2019diskann,krishnaswamy2024diskann} is Microsoft's widely used library for in-memory and SSD-backed search using the Vamana proximity graph.
Its SSD-backed design enables billion-scale search on a single machine by keeping compressed vectors in memory and fetching graph neighborhoods and full-precision vectors from disk.

\myparagraph{ParlayANN}
ParlayANN~\cite{manohar2024parlayann,parlayann_repo} is a highly optimized parallel graph-based ANNS library built on the ParlayLib~\cite{blelloch2020parlaylib} runtime.
It provides deterministic implementations of proximity-graph algorithms, including Vamana, designed to avoid synchronization bottlenecks and scale to large datasets on shared-memory machines.

\myparagraph{SVS}
Intel's Scalable Vector Search (SVS)~\cite{aguerrebere2023similarity} combines Vamana-based search with highly optimized SIMD kernels and compressed vector representations.
Its LVQ codec uses per-vector scaling and scalar quantization to reduce memory traffic, while LeanVec~\cite{tepper2024leanvec} combines dimensionality reduction with quantization.

\myparagraph{Hnswlib}
Hnswlib~\cite{hnswlib} provides a lightweight, header-only C++ implementation of HNSW~\cite{malkov2018efficient} with Python bindings and support for custom distance functions.
It supports incremental index construction, vector updates and deletions, and multithreaded search, with configurable graph connectivity and search effort.

\myparagraph{SymphonyQG}
SymphonyQG~\cite{gou2025symphonyqg,gao2025rabitq} combines RaBitQ quantization~\cite{gao2024rabitq} with FastScan~\cite{andre2016cache}, storing codes on graph edges to score neighbors sequentially.
Its layout aligns vertex degrees with 32-edge FastScan batches, trading increased memory usage from replicated codes for fewer random accesses during graph search.

\myparagraph{Faiss}
Faiss~\cite{johnson2019billion,douze2024faiss} is Meta's similarity search library, offering flat, IVF, and graph indexes with scalar and product quantization.
Its widely-used CPU implementations support large-scale search, with index and compression choices that provide different tradeoffs between memory usage, query throughput, and recall.

\myparagraph{ScaNN}
ScaNN~\cite{avq_2020,soar_2023} is an existing IVF-based similarity search library developed at Google, combining partitioning with anisotropic vector quantization and optimized scoring kernels for maximum inner-product search.
Its two-stage pipeline scores candidates using low-bit representations before reranking a smaller set, reducing the cost of evaluating large candidate pools.

\section{Nested TurboQuant (NTQ)}\label{sec:quant}\label{sec:quantization}

All production-grade ANNS libraries use \emph{quantization}. Because distance evaluation performs only a small constant number of arithmetic operations per loaded byte, ANNS is limited by memory bandwidth.
Compressing coordinates to low-precision codes reduces both memory traffic and the memory footprint of the index.

Graph beam search imposes a specific access pattern on the quantizer: each expansion fetches an arbitrary subset of non-contiguous points in $\mX$ and scores each point by scanning across its coordinates.
Unfortunately, standard PQ~\cite{jegou2011product,ge2013optimized}, as used in IVF indexes, is ill-suited for this task because loading each of the $M$ subspace lookup tables is expensive relative to scoring a single point.
Uniform scalar quantization methods avoid per-subspace codebooks and support fast coordinate-wise scans, but suffer higher quantization error.
Recently, rotation-based quantizers~\cite{vargaftik2021drive,vargaftik2022eden,gao2024rabitq,gao2025extendedrabitq,zandieh2025turboquant} have become popular.
They keep the fast coordinate-wise scans of uniform scalar quantizers and need no per-subspace tables, while achieving much lower quantization error.
A random rotation makes every coordinate approximately Gaussian, so a single scalar codebook serves all coordinates.
EDEN~\cite{vargaftik2022eden} and TurboQuant~\cite{zandieh2025turboquant} place this codebook non-uniformly at the Lloyd--Max points of the Gaussian~\cite{max1960quantizing,lloyd1982least}, which minimizes the mean-squared error per coordinate.

Graph search needs two precisions with conflicting goals.
Beam search scores thousands of random candidates per query and needs a compact code that is cheap to load from memory and score.
Reranking scores only the final beam, but needs the accuracy of floats to achieve very high recall.
A standard recipe is therefore to navigate with aggressive quantization and then rerank the converged beam with floats or higher-precision codes.
Storing a second code for reranking, or maintaining the float vectors, wastes significant memory.
A single nested code avoids this waste, but only if nesting does not degrade the quality of the codes.
We developed Nested TurboQuant (NTQ) to achieve this.

\myparagraph{Overview of Our Approach}
NTQ builds on TurboQuant~\cite{zandieh2025turboquant} and EDEN~\cite{vargaftik2022eden,benbasat2026eden}.
It takes the 4-bit Lloyd--Max codebook and splits each of its intervals into $2^k$ finer sub-intervals, which yields a $(4+k)$-bit code.
The 4-bit prefix of every code is therefore a complete TurboQuant code for navigation.
The full code nearly matches float accuracy for reranking.
Search cost stays fixed, while the $k$ additional bits per coordinate trade memory for reranking accuracy.
We mostly focus on 10- and 12-bit codes ($k = 6$ and $8$).
Nesting is nearly free, because the rotation makes every coordinate approximately Gaussian, and Gaussian sources are successively refinable~\cite{equitz1991successive}.
Both levels of NTQ are therefore close to optimal for the rotated coordinates.
We make both levels fast to score by evaluating their codebooks in registers rather than memory.

\textbf{The following subsections describe NTQ in detail for interested readers.}
Section~\ref{sec:quant-encoding} sets up the TurboQuant encoding that NTQ builds on and points out where we deviate from it.
Section~\ref{sec:quant-nested} introduces the nested refinement, which is the core of NTQ.
Sections~\ref{sec:quant-4bit} and~\ref{sec:quant-rerank} describe fast scoring kernels for 4-bit navigation and $(4+k)$-bit reranking.
All quantization and scoring kernels are implemented in portable SIMD using Highway~\cite{highway}.
For concreteness, we refer to AVX-512 instructions in the text.

\subsection{Rotation and Codebook}\label{sec:quant-encoding}

To encode a non-zero vector $x \in \mathbb{R}^d$, NTQ normalizes and rotates it to $y = \sqrt{d}\, R \hat{x}$ with $\hat{x} = x / \|x\|_2$.
It then quantizes each coordinate $y_i$ with a scalar codebook and stores a per-vector scale factor that corrects for the norm and for the shrinkage caused by quantization.

\myparagraph{Non-Power-of-Two Rotation}
Applying a dense orthogonal matrix $R$ costs $O(d^2)$ operations per vector, which is too expensive.
Therefore, we use a randomized Walsh--Hadamard transform~\cite{ailon2009fast,andoni2015practical}, which takes $O(d \log d)$ operations.
However, this transform is limited to dimensions that are a power of two.
Zero-padding $d$ to the next power of two as in EDEN~\cite{vargaftik2022eden} can nearly double the memory footprint and scoring work.

Instead, we factor $d = r \cdot 2^\ell$ with $r$ odd and $\ell \ge 0$ maximal.
Let $S = \operatorname{diag}(\sigma_1, \dots, \sigma_d)$ with independent uniform random signs $\sigma_i \in \{\pm 1\}$.
We split $Sx$ into $r$ contiguous blocks of $2^\ell$ coordinates and stack these blocks as the rows of a matrix $X \in \mathbb{R}^{r \times 2^\ell}$.
We then define $Rx$ as the concatenation of the rows of $Q X H_{2^\ell}$.
Here $H_{2^\ell} \in \{\pm 2^{-\ell/2}\}^{2^\ell \times 2^\ell}$ is the normalized (orthonormal) Walsh--Hadamard matrix, so right-multiplication applies the Walsh--Hadamard transform to each block.
The matrix $Q$ is an $r \times r$ Haar-random orthogonal matrix, so left-multiplication mixes corresponding coordinates across $r$ blocks.
Since $S$, $H_{2^\ell}$, and $Q$ are orthogonal, $R$ is an orthogonal matrix.
Evaluating $R x$ takes $O(d(\ell + r))$ operations, which is $O(d \log d)$ for small constant $r$ (\eg $r = 3$ for common dimensions $d \in \{96, 768, 1536, 3072\}$).
For these dimensions, zero-padding to the next power of two would add $33\%$ more codes and distance scoring work.

\myparagraph{Gaussian Lloyd--Max Codebook}
Following TurboQuant~\cite{zandieh2025turboquant}, we quantize each rotated coordinate using a symmetric $b$-bit Gaussian Lloyd--Max codebook~\cite{max1960quantizing,lloyd1982least}.
The codebook partitions $\mathbb{R}_{\ge 0}$ using $2^{b-1}$ positive centroids $0 < c_0 < \dots < c_{2^{b-1}-1}$ separated by midpoint thresholds $t_1, \dots, t_{2^{b-1}-1}$.
Each coordinate $y_i$ stores its sign $\operatorname{sgn}(y_i) \in \{\pm 1\}$ and the index $u_i \in \{0, \dots, 2^{b-1}-1\}$ of the centroid nearest to $|y_i|$, giving reconstruction $\tilde{y}_i = \operatorname{sgn}(y_i)\, c_{u_i}$.
NTQ uses this codebook with $b = 4$ for navigation, and Section~\ref{sec:quant-nested} refines its intervals for reranking.

\myparagraph{Norm and Shrinkage Rescaling}
Inner products with quantized vectors are biased unless they are corrected.
Under the standard normal density $\phi(z) = (2\pi)^{-1/2} e^{-z^2/2}$, the $b$-bit codebook incurs a per-coordinate mean-squared error
\begin{equation*}
D_b = \mathbb{E}[(y_i - \tilde{y}_i)^2] = 2 \sum_{u=0}^{2^{b-1}-1} \int_{t_u}^{t_{u+1}} (z - c_u)^2 \phi(z)\, dz,
\end{equation*}
with $t_0 = 0$ and $t_{2^{b-1}} = \infty$.
Because each centroid $c_u$ is the conditional mean of its interval, $\mathbb{E}[(y_i - \tilde{y}_i)\tilde{y}_i] = 0$, which gives $\mathbb{E}[y_i \tilde{y}_i] = \mathbb{E}[\tilde{y}_i^2] = 1 - D_b$.
Normalizing $\tilde{y}$ to unit length therefore shrinks inner products with $R \hat{x}$ by $\mathbb{E}[\langle R \hat{x}, \tilde{y} \rangle] / \sqrt{\mathbb{E}[\|\tilde{y}\|_2^2]} = \sqrt{1 - D_b}$.
Rotation-based quantizers correct this bias in different ways.
EDEN~\cite{vargaftik2022eden} divides by $\langle R x, \tilde{y} \rangle$.
TurboQuant~\cite{zandieh2025turboquant} uses $(b-1)$-bit Lloyd--Max quantization and reserves 1 bit per coordinate for a quantized Johnson--Lindenstrauss (QJL) sign sketch of the quantization residual.
We store the per-vector scale factor $s_b(x) = \|x\|_2 / \allowbreak (\sqrt{1 - D_b}\,\|\tilde{y}\|_2)$, so that $s_b(x)\,\tilde{y}$ estimates $R x$ without shrinkage bias.
This costs a single float per vector and leaves all $b$ bits to the codebook.
By orthogonality of $R$, we then estimate $\langle q, x \rangle = \langle R q, R x \rangle$ as $s_b(q)\, s_b(x)\, \langle \tilde{q}, \tilde{y} \rangle$ for a quantized query or $s_b(x)\, \langle R q, \tilde{y} \rangle$ for an unquantized query.

\subsection{Nested Refinement}\label{sec:quant-nested}

\begin{figure}[t]
  \centering
  \includegraphics[width=\columnwidth]{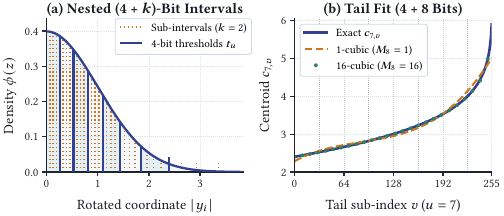}
  \caption{(a) Nested $(4+k)$-bit intervals under the rotated coordinate density $\phi(z)$, where each positive 4-bit interval $[t_u, t_{u+1})$ splits into $2^k$ sub-intervals without crossing $t_1, \dots, t_7$. (b) On the unbounded tail interval $u = 7$, a single cubic poorly fits the $(4+8)$-bit centroids $c_{7,v}$, whereas $M_8 = 16$ cubic segments track $c_{7,v}$ closely across all $256$ sub-indices.}
  \label{fig:nested-cubic}
\end{figure}

For reranking, NTQ adds $k$ refinement bits per coordinate to the 4-bit code.
A standard way to add such bits is two-stage residual quantization, which encodes the residual $y_i - \tilde{y}_i$ of the 4-bit reconstruction with a second codebook shared across all 4-bit intervals~\cite{jegou2011product,douze2024faiss}.
However, because the conditional distribution of a rotated coordinate varies significantly between the 4-bit intervals, a shared residual codebook is suboptimal.
In our experiments, residual quantization achieves lower recall than the nested codebook described next (see Section~\ref{sec:quant-ablations}).

We instead split each positive 4-bit interval $[t_u, t_{u+1})$ into $2^k$ contiguous sub-intervals with centroids $c_{u,v}$ for $v \in \{0, \dots, 2^k - 1\}$ (Figure~\ref{fig:nested-cubic}a).
Here $u \in \{0, \dots, 7\}$, $t_0 = 0$, and $t_8 = \infty$.
The centroids $c_{u,0}, \dots, c_{u,2^k-1}$ form a $2^k$-entry Lloyd--Max codebook for the standard normal density restricted to $[t_u, t_{u+1})$.
Each 4-bit interval therefore gets its own refinement codebook, adapted to the shape of $\phi(z)$ on that interval.
Encoding first finds the 4-bit interval $u_i$ of $|y_i|$ as in Section~\ref{sec:quant-encoding} and then the nearest centroid $c_{u_i, v_i}$ within that interval.
Since the sub-intervals never cross the 4-bit thresholds, each coordinate's $(4+k)$-bit code splits into its 4-bit code $u_i$ and a $k$-bit refinement code $v_i$.
The $(4+k)$-bit reconstruction is $\tilde{y}^{(4+k)}_i = \operatorname{sgn}(y_i)\, c_{u_i, v_i}$.
Its scale factor $s_{4+k}(x)$ follows Section~\ref{sec:quant-encoding}, with the mean-squared error $D_{4+k}$ summed over all $8 \cdot 2^k$ positive sub-intervals.

\myparagraph{Why Nesting Is Nearly Free}
Forcing the $(4+k)$-bit code to refine the 4-bit intervals could cost accuracy, because an unconstrained $(4+k)$-bit Lloyd--Max codebook would place all of its thresholds freely.
This constraint costs almost nothing, because Gaussian sources are successively refinable under mean-squared error~\cite{equitz1991successive}.

\myparagraph{Storage Layout}
The 4-bit codes, packed two per byte, and their per-vector scale factor form the navigation array read by beam search.
The refinement codes and $s_{4+k}(x)$ form a separate array read only for reranking, so beam search loads no extra bytes.

\subsection{Vectorized 4-Bit Scoring for Beam Search}\label{sec:quant-4bit}

Each coordinate of a data vector is stored as a 4-bit code for beam search, so the codebook has only 16 entries.
To score candidates with fast 8-bit integer dot-product instructions, we round these 16 centroids to int8 values and quantize the query coordinates to int8 as well.
Each codebook entry then occupies a single byte, so the entire codebook forms a 16-entry lookup table.
We load this table into a vector register.
Scoring a candidate looks up each of its 4-bit codes in this register and accumulates the products with the query coordinates.

\myparagraph{Codes and Table Entries}
For each database vector $x$, coordinate $i \in \{0, \dots, d-1\}$ packs its sign and 3-bit index $u_i \in \{0, \dots, 7\}$ into a 4-bit nibble $n_i \in \{0, \dots, 15\}$, which serves directly as the table index.
The table holds the 16 signed centroids $\pm c_0, \dots, \pm c_7$ quantized to int8 as $\pm\operatorname{round}(127\, c_u / c_7)$, so looking up $n_i$ yields $\bar{y}_i \in [-127, 127]$.
For a query $q$, we quantize the rotated coordinates to int8 as $\bar{q}_i = \operatorname{round}(127\, (Rq)_i / \|Rq\|_\infty) \in [-127, 127]$.
The scale factors $\bar{s}_4(x) = \|x\|_2 / (\sqrt{1 - D_4}\,\|\bar{y}\|_2)$ and $\bar{s}(q) = \|q\|_2 / \|\bar{q}\|_2$ absorb the int8 scales, so $\bar{s}(q)\,\bar{s}_4(x)\,\langle \bar{y}, \bar{q} \rangle$ estimates $\langle q, x \rangle$.

\myparagraph{Shifted Table Entries}
We accumulate the products with the x86 8-bit integer dot-product instruction \texttt{vpdpbusd}~\cite{intel_vnni}, which multiplies unsigned by signed bytes and accumulates into 32-bit integers.
The table therefore stores the shifted unsigned entries $\bar{y}_i + 128 \in \{1, \dots, 255\}$, while the query bytes $\bar{q}_i$ remain signed.
The shift adds a query-dependent, candidate-independent bias:
\begin{equation}\label{eq:bias-removal}
\sum_{i=0}^{d-1} (\bar{y}_i + 128)\,\bar{q}_i = \langle \bar{y}, \bar{q} \rangle + 128 \sum_{i=0}^{d-1} \bar{q}_i.
\end{equation}
We precompute this bias per query and subtract it from each candidate's accumulated dot product before multiplying by $\bar{s}(q)\,\bar{s}_4(x)$.

\myparagraph{Table Lookups}
The byte shuffle instruction \texttt{pshufb} performs the lookups.
It replaces each index byte in $\{0, \dots, 15\}$ by the table entry at that position.
With a copy of the table in each 128-bit lane, one AVX-512 shuffle performs 64 lookups.
For each candidate vector, coordinate pairs $(n_{2j}, n_{2j+1})$ are packed into the low and high nibbles of byte $j \in \{0, \dots, \lceil d/2 \rceil - 1\}$.
Rather than unpacking nibbles into coordinate order via cross-lane shuffles, we split $\bar{q}$ into contiguous even coordinates $(\bar{q}_0, \bar{q}_2, \dots)$ and odd coordinates $(\bar{q}_1, \bar{q}_3, \dots)$.
Each SIMD step loads one vector register of packed code bytes.
A bitwise mask (\texttt{0x0F}) extracts the even indices, and a 4-bit right shift followed by the same mask extracts the odd indices.
Two shuffles, each performing 64 table lookups, map both index streams to $\bar{y}_i + 128$, and two \texttt{vpdpbusd} instructions accumulate their dot products with the even and odd query coordinates.
Since each step takes only a few SIMD instructions, scoring is dominated by fetching neighbors from dispersed memory addresses.
We therefore issue software prefetches several candidates ahead during neighbor-list expansion to hide DRAM latency.

\subsection{Fast \texorpdfstring{$(4+k)$}{(4+k)}-Bit Reranking}\label{sec:quant-rerank}

Reranking fetches the refinement codes of the top candidates.
It scores each candidate against the unquantized rotated query $R q$ as $s_{4+k}(x)\, \langle R q, \tilde{y}^{(4+k)} \rangle$.
Unlike the int8 table of Section~\ref{sec:quant-4bit}, reranking computes in float precision.
Refinement codebooks for small $k$ fit in registers, but the 10- and 12-bit codebooks have $512$ or $2048$ positive centroids and need memory gathers, which take roughly $80\%$ of reranking time at $d = 4096$.

\myparagraph{Piecewise-Cubic Codebook Approximation}
We therefore replace the table with a piecewise-cubic polynomial that computes each centroid $\hat{c}_{u,v}$ in registers from its integer index.
The coefficients of each piece are fit offline by least squares to the codebook centroids.

On each of the seven bounded intervals ($u \le 6$), $\phi(z)$ is smooth and bounded away from zero.
A single cubic per interval therefore fits all $2^k$ sub-centroids with small error.
On the unbounded tail $[t_7, \infty)$, $\phi(z)$ decays rapidly and the sub-intervals grow progressively wider, so a single cubic cannot fit all tail centroids (Figure~\ref{fig:nested-cubic}b).
We split the $2^k$ tail sub-intervals into $M_k$ equal segments of $B_k = 2^k / M_k$ sub-intervals each.
$M_k$ must be a power of two so that $B_k$ is one as well, which lets each SIMD lane find its segment with a shift and a mask.
We choose $M_6 = 8$ and $M_8 = 16$, the largest such values for which the coefficients of all $7 + M_k$ segments fit in one 16-lane float register per polynomial degree for $k = 6$, and in two for $k = 8$.

\myparagraph{In-Register SIMD Evaluation}
A coordinate with codes $(u, v)$ belongs to segment $s = u$ with local index $\tau = v$ if $u < 7$.
In the tail, it belongs to segment $s = 7 + \lfloor v / B_k \rfloor$ with $\tau = v \bmod B_k$.
We evaluate 16 coordinates in parallel, and each SIMD lane computes $(s, \tau)$ with a shift and a mask.
Each lane then selects its segment's coefficients $(a_{s,0}, \dots, a_{s,3})$ with an in-register permutation (\texttt{vpermps} for $k = 6$, or two-register \texttt{vpermi2ps} for $k = 8$).
Finally, it evaluates the cubic in Horner form with three fused multiply-adds:
\begin{equation}\label{eq:piecewise-cubic}
\hat{c}_{u,v} = a_{s,0} + \tau \bigl(a_{s,1} + \tau (a_{s,2} + \tau a_{s,3})\bigr).
\end{equation}
Replacing memory gathers with in-register permutations and fused multiply-adds roughly halves 10- and 12-bit reranking latency, with no loss in recall.

\section{Search Optimizations}
\label{sec:search}
Although the basic beam search (Algorithm~\ref{alg:beam_search}) is simple, ANN systems vary significantly in the details of how the search is implemented.
In \sys{}, we build on our fast 4-bit scoring kernel (Section~\ref{sec:quant-4bit}) with search optimizations that reduce memory traffic and the overhead of maintaining the candidate beam.

\subsection{Search on Projected Coordinates}

On high-dimensional datasets, fetching and scoring even quantized candidate vectors consumes substantial memory bandwidth.
In \sys{}, we reduce this cost by projecting database vectors and queries into a lower-dimensional space using Principal Component Analysis (PCA)~\cite{jolliffe2016principal}.
We encode the projected vectors with 4-bit navigation codes and score them using the kernel from Section~\ref{sec:quant-4bit}.
Each candidate evaluation therefore reads and scores fewer coordinates, reducing memory traffic beyond quantization alone.

The database vectors can be projected offline, whereas each incoming query must be multiplied by the projection matrix $P$ before navigation begins.
This matrix multiplication is itself memory-bound and competes with graph traversal for memory bandwidth when queries run in parallel across threads.
To alleviate this bottleneck, we also quantize the \emph{rows} of $P$ to 4 bits, reducing the data fetched during query projection.
Our results show that this does not impair recall (Figure~\ref{fig:recall_qps_pca_matrix_repr_3ds}).

\subsection{Candidate Prefiltering}\label{sec:search-prefilter}

During beam search, most scanned neighbors $v \in N(p)$ are not useful and will never enter the beam $B$.
On high-dimensional datasets, we can use a cheaper filter distance $d_{\mathrm{filt}}(q, v)$ to prune unpromising candidates early and thus avoid computing the 4-bit navigation distance $d_{\mathrm{nav}}(q, v)$.
Only the neighbors with $d_{\mathrm{filt}}(q, v)$ below a cutoff threshold are scored with $d_{\mathrm{nav}}$.

This general idea has been used in several recent systems~\cite{chen2023finger,gao2023adsampling,lu2024probabilistic,Angular25,lu2026probabilistic,sun2026fastgann}.
For example, ADSampling~\cite{gao2023adsampling} uses Johnson--Lindenstrauss projections for Euclidean distance, and FINGER~\cite{chen2023finger} uses angle approximation.
However, prior methods rely on a specific relationship between the two distance measures so that a candidate's filter score can be compared against the navigation distance $d_{\mathrm{nav}}(q, b_L)$ of the furthest point $b_L \in B$~\cite{lu2025effective}.

\myparagraph{Our Approach} We instead allow pairing \emph{arbitrary} filter and navigation measures, such as Hamming distance and inner product.
All we require is that each measure independently approximates the distance ordering relative to $q$, which makes this framework flexible.
This decoupling works because beam admission depends only on relative distance ordering around $q$ rather than calibrated distance values.
Any coarse sketch whose ranking correlates with $d_{\mathrm{nav}}(q, \cdot)$ can separate the few promising neighbors from the many that lie far behind the beam boundary.
At the same time, $d_{\mathrm{filt}}$ has higher variance than $d_{\mathrm{nav}}$ and cannot replace it for graph navigation, even in early hops.
Navigating directly with $d_{\mathrm{filt}}$ commits the greedy traversal to its ranking errors and can steer the beam away from the query neighborhood.
Using $d_{\mathrm{filt}}$ solely as a conservative gate backed by $d_{\mathrm{nav}}$ preserves the 4-bit search trajectory while skipping full distance evaluations on clearly unpromising neighbors.
Because the two measures may not be comparable, we derive the filter cutoff from the beam using $d_{\mathrm{filt}}$ itself.
Thresholding against $d_{\mathrm{filt}}(q, b_L)$ alone would be brittle, because a noisy filter can assign $b_L$ a small distance by chance and prune candidates that beat $b_L$ under $d_{\mathrm{nav}}$.
While the beam is filling up ($|B| < L$), prefiltering is disabled.
Once the beam is full (at step $t_0$), we compute the mean filter distance $\mu_t$ over the $w$ furthest points at the tail of $B_t = (b_1, \dots, b_L)$ to smooth out single-point noise.
We then set the cutoff $\tau_t$ to the running average of $\mu_s$ over all steps since $t_0$:
\begin{equation}\label{eq:running-threshold}
\mu_t = \frac{1}{w} \sum_{j = L - w + 1}^{L} d_{\mathrm{filt}}(q, b_j), \qquad
\tau_t = \frac{1}{t - t_0 + 1} \sum_{s = t_0}^{t} \mu_s.
\end{equation}
If $b_L$ did not change on step $t$, we reuse $\mu_t = \mu_{t-1}$.
As the beam improves, $\mu_t$ decreases, so $\tau_t$ tightens over time while staying slightly looser than $\mu_t$.
We instantiate this scheme with two filters:

\myparagraph{TurboSlice}
Because the random rotation in Section~\ref{sec:quant-encoding} spreads a vector's energy evenly across all rotated coordinates, any prefix of $d_{\mathrm{ts}} < d$ rotated 4-bit coordinates acts as a low-dimensional sketch, which provides an unbiased estimator of the dot product~\cite{daliri2026turboslice}.
This sketch sits at the front of the 4-bit navigation code and requires no extra vector storage.
For typical prefix lengths $d_{\mathrm{ts}} \in \{128, 256\}$, the filter reads only one or two cache lines and skips loading the rest of the vector for rejected neighbors.
We score the prefix with the 4-bit kernel of Section~\ref{sec:quant-4bit} over the first $d_{\mathrm{ts}}$ coordinates and rescale by a precomputed prefix norm factor $\bar{s}_4^{(d_{\mathrm{ts}})}(x) = \|x\|_2 / (\sqrt{1 - D_4}\,\|\bar{y}_{1:d_{\mathrm{ts}}}\|_2)$.

\myparagraph{1-Bit Sparse Johnson--Lindenstrauss (JL)}
ParlayANN~\cite{manohar2024parlayann} already offers a filter based on JL, whose details have not been published so far, however. 
This filter maps each vector $x \in \mathbb{R}^d$ to a $d_{\mathrm{jl}}$-bit sign sketch $h(x) \in \{0, 1\}^{d_{\mathrm{jl}}}$ with bits $h_i(x) = \mathbf{1}[\langle a_i, x \rangle > 0]$ (we usually use $d_{\mathrm{jl}} = 1024$).
We pack the sketches into 64-bit words and compare them by Hamming distance $\|h(q) \oplus h(v)\|_1$.
Because our cutoff~\eqref{eq:running-threshold} only compares filter scores against other filter scores, we can use the Hamming distances directly without reconstructing angles or norms.
To make projecting the query fast, it suffices to use a sparse random projection matrix $A \in \{-1, 0, +1\}^{d_{\mathrm{jl}} \times d}$ with only a few non-zero entries per row.

While our experiments show that PCA is generally more effective than the prefilters and they do not stack well (both reduce memory traffic in a similar way), we note that the prefilters are quite useful in scenarios where we cannot work with PCA.
One example is network-constrained deployments, where we do not want to send multiple sets of vectors per query over the wire.
Here, taking a TurboSlice prefix does not incur additional traffic.
Another example is extremely high-dimensional vectors, where the projection matrix is so large that it slows down projection.

\subsection{Beam Management}\label{sec:search-beam}

A naive beam search maintains two priority queues for the closest unexplored vertices and the furthest points in $B$.
However, priority-queue operations take $O(\log L)$ time and have poor cache locality.
Following ParlayANN~\cite{manohar2024parlayann} and SVS~\cite{aguerrebere2023similarity}, we instead keep $B$ in a single array of size at most $L$ sorted by distance.
Each entry stores a flag indicating whether it has been explored, and we track the index of the first unexplored entry.
Picking the next vertex to expand then takes $O(1)$ time.
Because $B$ is sorted, the furthest element $b_L$ sits at the end of the array and provides the distance cutoff for entering the beam.
Inserting new points into the sorted array one at a time would take $O(L)$ time per point.
Instead, when expanding a vertex $p \in B$, we keep only the unvisited neighbors that beat $b_L$, sort the survivors, and merge them into $B$ in a single linear pass.

Later in the search, most neighbors have already been visited or fail to beat $b_L$, so merging after every expansion scans the whole beam just to insert one or two points.
Once the initial phase of the search is over, we therefore batch across consecutive expansions.
We collect unvisited neighbors from several unexplored beam vertices, score them together, and merge the survivors into $B$ in a single pass.
Batching also exposes a longer stream of candidates, which helps amortize startup latency (due to prefetching) in the TurboQuant scoring kernel.

\section{Graph Compression}
\label{sec:compress}
\providecommand{\sys}{\textsc{Ginger}}

In \sys{}, we also reduce the memory occupied by the proximity graph itself, which can exceed the footprint of the quantized vectors on low-dimensional datasets.
For example, a 96-dimensional vector occupies 144 bytes at 12 bits per coordinate, whereas a flat adjacency list with 64 four-byte slots and a degree counter occupies 260 bytes.
We therefore compress the graph losslessly while keeping decoding fast enough to avoid slowing search.

Our approach first relabels vertices so that neighbors receive nearby IDs (Section~\ref{sec:compress:reordering}), then stores each sorted neighbor list as gaps between consecutive IDs.
We encode these gaps with a codec that packs small values into fewer than eight bits and decodes four at a time using a few SIMD instructions (Section~\ref{sec:compress:codecs}).

\myparagraph{Gap encoding}
Let $v_1 < v_2 < \dots < v_{\deg(u)}$ be the sorted neighbor IDs of a vertex $u$.
Gap encoding stores the first ID $g_1 = v_1$, followed by the differences $g_i = v_i - v_{i-1}$ for $2 \le i \le \deg(u)$.
A gap $g_i$ needs about $\log_2(g_i + 1)$ bits, so small gaps mean a small graph.
A prefix sum over the gaps recovers the IDs.

\subsection{Graph Reordering}\label{sec:compress:reordering}

To reduce gaps between consecutive neighbor IDs, we place well-connected groups of vertices into contiguous ID ranges, an approach well studied for web and social graphs~\cite{randall2002link,blandford2003compact,boldi2004webgraph,chierichetti2009compressing,apostolico2009graph,boldi2011layered,safro2011multiscale,lim2014slashburn,shun2015smaller} and inverted indexes~\cite{blandford2002index,shieh2003inverted,blanco2005document,silvestri2007sorting,ding2010scalable,dhulipala2016compressing}.
In \sys{}, recursive graph partitioning (\textsf{RGP}) uses a parallel multilevel partitioner~\cite{gottesbueren2021deep,gottesburen2024unleashing} to recursively split the graph into balanced blocks with few crossing edges.
We number the small leaf blocks in recursion order, assign consecutive IDs within each block, and permute the vectors accordingly.

Balanced graph partitioning only works on undirected graphs, hence we symmetrize the graph by adding the reverse of every edge.
Reciprocal connections then carry twice the weight, reflecting the potential for long gaps in both adjacency lists when their endpoints are separated.
\textsf{RGP} yields smaller gaps than recursive graph bisection~\cite{dhulipala2016compressing} on all six datasets and is faster to compute on all but the smallest (Section~\ref{sec:compression-ablations}).

\subsection{AdaptiveGap: Sub-Byte Gap Coding with SIMD Decoding}\label{sec:compress:codecs}

Beam search decodes a neighbor list at every expansion, so graph compression must keep decoding cheap enough to preserve query throughput.
In codecs such as PrefixVarint~\cite{dean2009challenges}, each gap takes one to five bytes and begins with a unary length prefix, so locating the next gap requires reading the current gap's length.

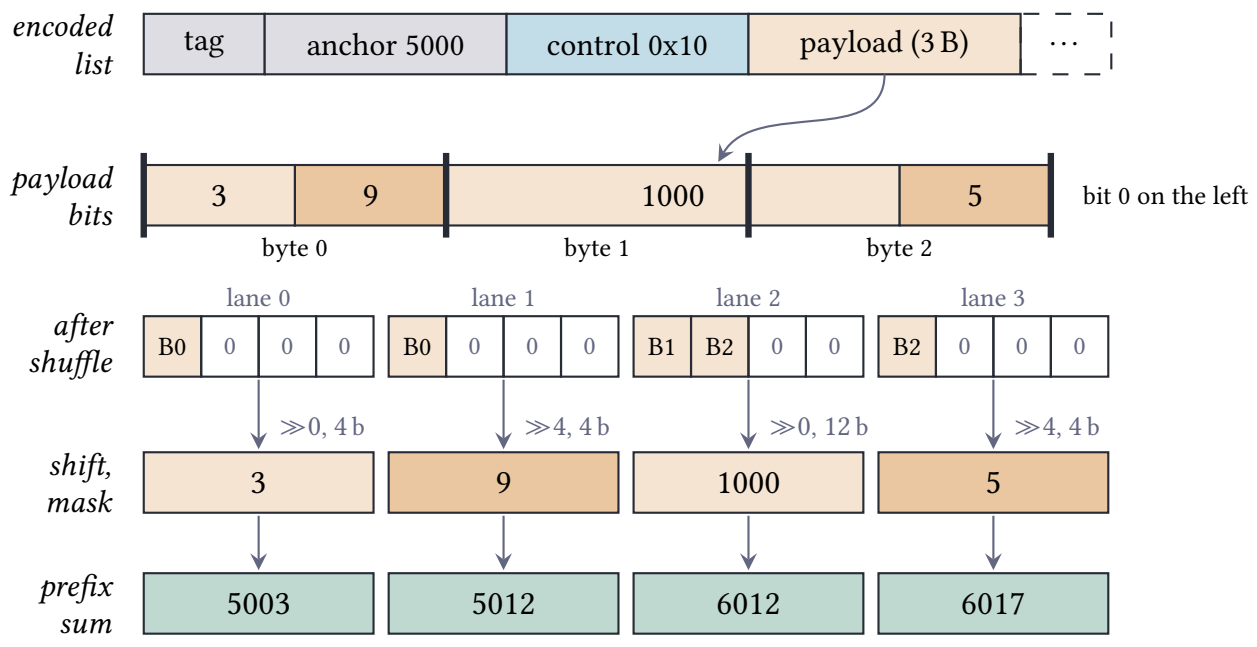
\begin{figure}[t]
  \centering
\definecolor{agSteel}{HTML}{636681}%
\definecolor{agAegean}{HTML}{519EBE}%
\definecolor{agRust}{HTML}{CC791E}%
\definecolor{agGinger}{HTML}{2C7F61}%
\definecolor{agInk}{HTML}{262B35}%
\colorlet{agHeader}{agSteel!22}%
\colorlet{agControl}{agAegean!35}%
\colorlet{agPayload}{agRust!20}%
\colorlet{agPayloadAlt}{agRust!42}%
\colorlet{agDecoded}{agGinger!30}%
\colorlet{agNote}{agSteel}%
\begin{tikzpicture}[
    x=1cm, y=1cm, font=\scriptsize, >=stealth,
    box/.style={draw=agInk, minimum height=4mm, inner sep=0pt, outer sep=0pt},
    lbl/.style={anchor=east, align=right, font=\scriptsize\itshape},
  ]
  \def\yA{0}
  \node[lbl] at (-0.1, \yA) {encoded\\list};
  \node[box, fill=agHeader, minimum width=0.8cm, anchor=west] (tag) at (0, \yA) {tag};
  \node[box, fill=agHeader, minimum width=1.6cm, anchor=west] (anc) at (tag.east) {anchor 5000};
  \node[box, fill=agControl, minimum width=1.6cm, anchor=west] (ctl) at (anc.east) {control 0x10};
  \node[box, fill=agPayload, minimum width=1.8cm, anchor=west] (pay) at (ctl.east) {payload (3\,B)};
  \node[box, dashed, minimum width=0.6cm, anchor=west] (more) at (pay.east) {\dots};

  \def\yB{-1.0}
  \node[lbl] at (-0.1, \yB) {payload\\bits};
  \foreach \x/\w/\v/\f in {0/4/3/agPayload, 4/4/9/agPayloadAlt, 8/12/1000/agPayload, 20/4/5/agPayloadAlt} {
    \draw[agInk, fill=\f] (0.25*\x, \yB-0.2) rectangle ++(0.25*\w, 0.4);
    \node at (0.25*\x+0.125*\w, \yB) {$\v$};
  }
  \foreach \b in {0, 1, 2} {
    \draw[agInk, very thick] (2*\b, \yB-0.28) -- ++(0, 0.56);
    \node[font=\tiny] at (2*\b+1, \yB-0.36) {byte \b};
  }
  \draw[agInk, very thick] (6, \yB-0.28) -- ++(0, 0.56);
  \node[font=\tiny, anchor=west] at (6.1, \yB) {bit 0 on the left};
  \draw[->, agNote] (pay.south) to[out=-90, in=60] (3.8, \yB+0.22);

  \def\yC{-2.0}
  \node[lbl] at (-0.1, \yC) {after\\shuffle};
  \foreach \l/\ba/\bb in {0/0/9, 1/0/9, 2/1/2, 3/2/9} {
    \fill[agPayload] (1.62*\l, \yC-0.2) rectangle ++(0.38, 0.4);
    \ifnum\bb<9 \fill[agPayload] (1.62*\l+0.38, \yC-0.2) rectangle ++(0.38, 0.4);\fi
    \foreach \s in {0,...,3} {
      \draw[agInk] (1.62*\l+0.38*\s, \yC-0.2) rectangle ++(0.38, 0.4);
    }
    \node[font=\tiny] at (1.62*\l+0.19, \yC) {B\ba};
    \ifnum\bb<9 \node[font=\tiny] at (1.62*\l+0.57, \yC) {B\bb};\fi
    \node[font=\tiny, agNote] at (1.62*\l+0.76, \yC+0.33) {lane \l};
  }
  \foreach \l/\s in {0/1, 0/2, 0/3, 1/1, 1/2, 1/3, 2/2, 2/3, 3/1, 3/2, 3/3} {
    \node[font=\tiny, agNote] at (1.62*\l+0.38*\s+0.19, \yC) {0};
  }

  \def\yD{-2.9}
  \node[lbl] at (-0.1, \yD) {shift,\\mask};
  \foreach \l/\sh/\w/\v/\f in {0/0/4/3/agPayload, 1/4/4/9/agPayloadAlt, 2/0/12/1000/agPayload, 3/4/4/5/agPayloadAlt} {
    \draw[->, agNote] (1.62*\l+0.76, \yC-0.22) -- (1.62*\l+0.76, \yD+0.22);
    \node[font=\tiny, agNote, anchor=west] at (1.62*\l+0.78, \yD+0.36) {${\gg}\sh$, $\w$\,b};
    \node[box, fill=\f, minimum width=1.52cm] at (1.62*\l+0.76, \yD) {$\v$};
  }

  \def\yE{-3.7}
  \node[lbl] at (-0.1, \yE) {prefix\\sum};
  \foreach \l/\v in {0/5003, 1/5012, 2/6012, 3/6017} {
    \draw[->, agNote] (1.62*\l+0.76, \yD-0.22) -- (1.62*\l+0.76, \yE+0.22);
    \node[box, fill=agDecoded, minimum width=1.52cm] at (1.62*\l+0.76, \yE) {$\v$};
  }
\end{tikzpicture}
  \caption{Encoding and SIMD decoding of an AdaptiveGap list in mode $\{4, 12, 20, 28\}$. With $5000$ stored as the anchor, the control byte selects widths $4, 4, 12, 4$ (2-bit selectors $0, 0, 1, 0$ packed from the least significant bit) for gaps $3, 9, 1000, 5$, which occupy three payload bytes. One byte shuffle followed by per-lane shifts and masks recovers the gaps, and a prefix sum reconstructs the neighbor IDs.}
  \label{fig:adaptivegap}
\end{figure}

\myparagraph{StreamVByte}
We build on StreamVByte~\cite{lemire2018streamvbyte}, which removes this dependency by storing lengths separately from gaps.
It encodes groups of four gaps using a \emph{control byte} with four 2-bit length selectors and a separate \emph{payload} holding the gap values back to back.
Each selector specifies a gap length of one to four bytes.
To decode a group, StreamVByte loads 16 payload bytes into a SIMD register and uses one byte shuffle (\texttt{pshufb}) to distribute the values among four 32-bit lanes.
A precomputed \emph{shuffle mask} for each of the 256 control bytes specifies which payload bytes to copy into each lane, and a SIMD prefix sum recovers the neighbor IDs.
However, rounding each gap to whole bytes wastes space; for example, a 9-bit gap requires two bytes.

\myparagraph{AdaptiveGap}
AdaptiveGap retains StreamVByte's single-shuffle decoding while adding adaptive bit widths and optional separate storage of the first neighbor ID.
Each neighbor list selects one of 16 \emph{modes}, each specifying four bit widths that the 2-bit selectors choose among.
Mode~0 uses the widths $\{8, 16, 24, 32\}$, which corresponds to StreamVByte, while the other modes use sub-byte widths, such as $\{4, 12, 20, 28\}$.
For example, StreamVByte stores the gaps $3$, $9$, $1000$, and $5$ in $1 + 1 + 2 + 1 = 5$ bytes.
In the mode $\{4, 12, 20, 28\}$, they occupy $4 + 4 + 12 + 4 = 24$ bits, or 3 bytes, with the same single control byte.

AdaptiveGap packs gaps without padding within each byte-aligned group, so adjacent gaps can share a byte.
In Figure~\ref{fig:adaptivegap}, for example, the gap 9 starts at bit 4 of byte 0.
A single shuffle gathers the bytes covering each gap into a 32-bit lane, and a right shift and mask isolate its value.
For this to work, all bytes that contain a gap must fit into its lane.
A gap of width $w$ that starts at bit offset $o$ of a byte spans $o + w$ bits, so every gap must satisfy $o + w \le 32$.
In mode~0, all gaps are byte-aligned, so the offset is always 0.
In the other modes, all widths are even, so the offset is at most 6, and widths up to 26 bits are safe.
If all widths of a mode are multiples of 4, the offset is 0 or 4, and widths up to 28 bits are safe.
Unlike in StreamVByte, the shuffle mask depends on both the mode and the control byte, so we precompute a mask, along with the per-lane shifts and bit masks, for each of the $16 \times 256$ combinations.

The first entry $g_1 = v_1$ is an absolute ID and can require 30 bits at $n = 10^9$, potentially forcing the entire list onto mode~0.
AdaptiveGap avoids this by optionally storing the first ID separately as a varint, called the \emph{anchor}.
The encoder picks the mode, with or without the anchor, that minimizes the encoded size, and a 1-byte tag at the start of the list records its choice.
As in a CSR layout, an offset array locates each encoded list, and each list stores its degree.

\myparagraph{Summary}
Together, reordering and AdaptiveGap shrink the graph by $2.1\times$ on average (up to $2.5\times$) relative to a CSR layout without slowing down queries (Section~\ref{sec:compression-ablations}).
Compared with StreamVByte on the same ordering, AdaptiveGap stores an average of $0.9$ fewer bits per edge (up to $1.2$ bits).
This reduction saves a further $5.4\%$ of graph memory (up to $7.6\%$).

\section{Experimental Evaluation}
\label{sec:eval}

\begin{figure*}[t]
  \centering
  \includegraphics[width=0.95\textwidth]{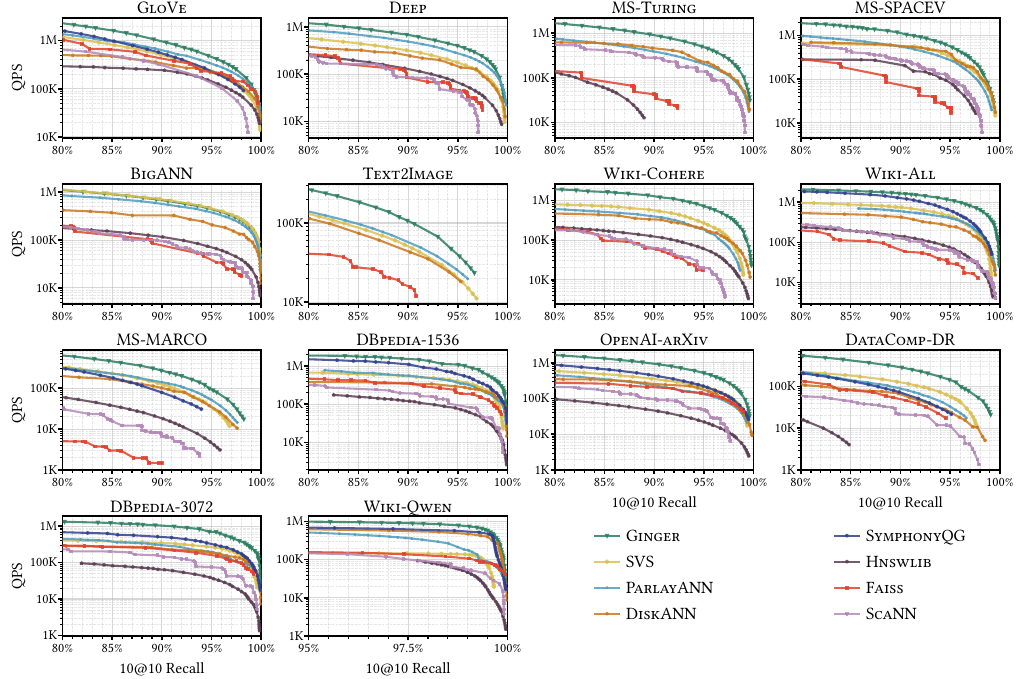}
  \caption{Throughput in queries per second (QPS) vs.\ recall $10@10$. \textsc{Ginger} uses 4+8-bit NTQ and graph compression. Missing curves are largely due to out-of-memory failures (see Table~\ref{tab:construction_times}). \textsc{SVS} on \textsc{MS-Turing} leaves the graph disconnected from the entry point. \textsc{SymphonyQG} on \textsc{Wiki-Cohere} reaches at most $\approx 0.5\%$ recall.
  }
  \label{fig:recall_qps}
\end{figure*}

\begin{table}[t]
  \footnotesize
  \setlength{\tabcolsep}{3.5pt}
  \caption{Overview of datasets used in our evaluation, where $d$ is the dimensionality of the vectors, $n$ is the number of vectors, and $n_q$ is the number of queries.
  }
  \label{tab:datasets}
  \centering
  \begin{tabular}{clccccc}
    \toprule
    & Dataset & $d$ & $n$ & $n_q$ & Type & Metric \\
    \midrule
    \multirow{6}{*}{\rotatebox[origin=c]{90}{Low dimensional}}
    & \textsc{GloVe}~\cite{pennington2014glove} & 100 & 1.2M & 10k & float & MIPS \\
    & \textsc{Deep}~\cite{babenko2016efficient} & 96 & 1B & 10k & float & L2 \\
    & \textsc{MS-Turing}~\cite{simhadri2022results} & 100 & 1B & 100k & float & L2 \\
    & \textsc{MS-SPACEV}~\cite{simhadri2022results} & 100 & 1B & 29k & int8 & L2 \\
    & \textsc{BigANN}~\cite{jegou2011product} & 128 & 1B & 10k & uint8 & L2 \\
    & \textsc{Text2Image}~\cite{simhadri2022results} & 200 & 1B & 100k & float & MIPS \\
    \midrule
    \multirow{8}{*}{\rotatebox[origin=c]{90}{High dimensional}}
    & \textsc{Wiki-Cohere}~\cite{reimers2022cohere} & 768 & 35M & 5k & float & MIPS \\
    & \textsc{Wiki-All}~\cite{rapids2024wikiall} & 768 & 88M & 10k & float & MIPS \\
    & \textsc{MS-MARCO}~\cite{simhadri2024results} & 768 & 101.07M & 9376 & float & MIPS \\
    & \textsc{DBpedia-1536}~\cite{qdrant2024dbpedia1536} & 1536 & 999k & 1k & float & L2 \\
    & \textsc{OpenAI-arXiv}~\cite{neelakantan2022text} & 1536 & 2.32M & 20k & float & L2 \\
    & \textsc{DataComp-DR}~\cite{apple2024datacompdr} & 1536 & 12.78M & 1k & float & L2 \\
    & \textsc{DBpedia-3072}~\cite{qdrant2024dbpedia3072} & 3072 & 999k & 1k & float & L2 \\
    & \textsc{Wiki-Qwen}~\cite{maknee2024wikipediaqwen8b} & 4096 & 500k & 100k & float & L2 \\
    \bottomrule
  \end{tabular}%
\end{table}

We compare \textsc{Ginger} with seven widely used ANNS libraries across 14 large-scale datasets ranging up to a billion points and up to four-thousand dimensions. We assess query throughput, memory footprint, and indexing time for each method.
Subsequently, we run ablation studies to isolate where our performance gains come from and how key design choices affect performance, including standalone evaluations of NTQ and AdaptiveGap (Section~\ref{sec:ablations}).

\subsection{Experimental Setup}

\myparagraph{Machine}
We use a Google Cloud \texttt{c4-\allowbreak highmem-\allowbreak 288} instance, a dual-socket server with two 72-core Intel Xeon 6985P-C processors (Granite Rapids) and 2.3\,TB of RAM across 6 NUMA nodes.

\myparagraph{Datasets}
We evaluate \textsc{Ginger} on 14 public ANNS datasets covering Euclidean and maximum inner-product search (MIPS), summarized in Table~\ref{tab:datasets}.
Six low-dimensional datasets ($d \leq 200$) comprise \textsc{GloVe} and five billion-vector benchmarks spanning image, web-search, and cross-modal retrieval.
Eight high-dimensional datasets ($768 \leq d \leq 4096$) contain up to 101 million vectors and cover text, encyclopedia, scientific, multimodal, and LLM embeddings.
Input coordinates are 8-bit integers for \textsc{MS-SPACEV} and \textsc{BigANN}, and 32-bit floats for the remaining twelve datasets.

\myparagraph{Existing Libraries}
We compare \sys{} against the seven libraries introduced in Section~\ref{sec:prior-libraries}: the Vamana implementations in ParlayANN~\cite{manohar2024parlayann,parlayann_repo}, SVS~\cite{aguerrebere2023similarity}, and DiskANN~\cite{jayaram2019diskann,krishnaswamy2024diskann}; HNSW~\cite{malkov2018efficient} from Hnswlib~\cite{hnswlib}; the IVF implementations in ScaNN~\cite{avq_2020,soar_2023} and Faiss~\cite{douze2024faiss}; and SymphonyQG~\cite{gou2025symphonyqg} from the RaBitQ-Library~\cite{gao2025rabitq}.
Together, these libraries form a comprehensive comparison set spanning both widely adopted industrial systems and highly optimized research implementations.
We select SymphonyQG to represent methods that store auxiliary data on graph edges because of its reported query performance and scalable, multithreaded C++ implementation~\cite{gou2025symphonyqg,gao2025rabitq}.

\myparagraph{Environment}
All C++ implementations are compiled with Clang v19.1.7.
DiskANN is built with Rust v1.97.0.
We use full compile-time (\texttt{-O3}) and link-time optimizations (\texttt{-lto}), and native architecture tuning (\texttt{-march=\allowbreak native}).
To avoid allocator contention and ensure a fair comparison, we link all implementations against \texttt{jemalloc} v5.3.0.
This change is either neutral or positive on QPS for all methods.
To ensure accurate query timings, we perform one warm-up run (discarded) followed by 5 timed runs of the query set for each parameter configuration and report the average throughput. To prevent cache interference across runs, we flush the CPU cache between runs.  

\myparagraph{Query Setup}
We evaluate query throughput without batching: individual queries are processed in parallel across threads. To obtain the Pareto-optimal frontier of throughput at each recall level, we sweep the search beam width $L_{\text{search}}$ with select values in the range $[10, 2000]$ for graph-based indexes.
For IVF-based indexes, we sweep the number of probed leaf clusters $n_{\text{probe}}$ over select values in $[1, 1000]$ and the number of candidates to rerank $n_{\text{rerank}} \in \{2k, 4k, 8k, 16k, 32k, 64k\}$. For multi-level \textsc{ScaNN}, we additionally sweep the number of top-level clusters to probe on two-level trees, and for three-level trees we evaluate selected pairs of middle- and top-level probe counts to cover various routing profiles that vary the search cost at each level, which we model by the amount of data loaded. See Appendix~\ref{sec:param-config} for more details.

Table~\ref{tab:query_configs} in Appendix~\ref{sec:param-config} summarizes the vector representations used by each library.
On the 8-bit integer datasets (\textsc{MS-SPACEV} and \textsc{BigANN}), all graph libraries operate directly on the native \textsf{int8}/\textsf{uint8} coordinates without quantization.
On high-dimensional float datasets ($d \geq 768$), libraries combine dimensionality reduction (PCA, OPQ, or JL prefiltering) or 1--2-bit spherical codes with higher-precision reranking.
On low-dimensional float datasets ($d \leq 200$), dimensionality reduction degrades accuracy so we omit it.
 
\myparagraph{Index Parameters}
All graph-based indexes are constructed with maximum out-degree $R = 64$ and construction beam width $L = 128$ (corresponds to $M = 32, ef_C = 128$ for \textsc{Hnswlib}).
\textsc{PiPNN} uses the standard parameters from the original paper: leaf size $C = 1024$ and tree fanout $(10, 3)$.
On \textsc{GloVe}, we use $R = 100, L = 250$ for \textsc{Vamana}, and $M = 50, ef_C = 250$ for \textsc{Hnswlib}.
For \textsc{Vamana} and \textsc{PiPNN} based methods we use the standard two build passes, except for \textsc{DiskANN} which only supports single-pass. We instead double the insertion beam-width to match the build effort and improve index quality.
For \textsc{ScaNN}, we use anisotropic vector quantization ($\text{AVQ} = 2.5$) and spilling with orthogonality-amplified residuals ($\text{SOAR} = 1.5$, with $\lambda = 1.0$ on \textsc{GloVe} and \textsc{Wiki-Cohere}, and $\lambda = 2.0$ on \textsc{DataComp-DR} and \textsc{DBpedia}).
We set the noise shaping threshold to $\text{NST} = 0.2$ for unit-norm datasets ($\text{NST} = 0.1$ on \textsc{OpenAI-arXiv}, \textsc{DataComp-DR}, and \textsc{DBpedia}), and scale it proportionally to the average vector norm on non-unit-norm datasets ($\text{NST} = 24.0$ on \textsc{MS-SPACEV}, $100.0$ on \textsc{BigANN}, $2.75$ on \textsc{Wiki-Cohere}, and $2.0$ on \textsc{MS-MARCO}).
Refer to Table~\ref{tab:construction_params} for the dataset-specific parameters such as RobustPrune $\alpha$ values, partition sizes, and IVF quantizer configurations.
All methods use multi-threading for the builds.

To support reproducibility, we will release our experimental setup, including the benchmark harness, build scripts, and dataset configurations alongside.

\begin{figure*}[t]
  \centering
  \includegraphics[width=0.9\textwidth, trim=0 7.9pt 0 0, clip]{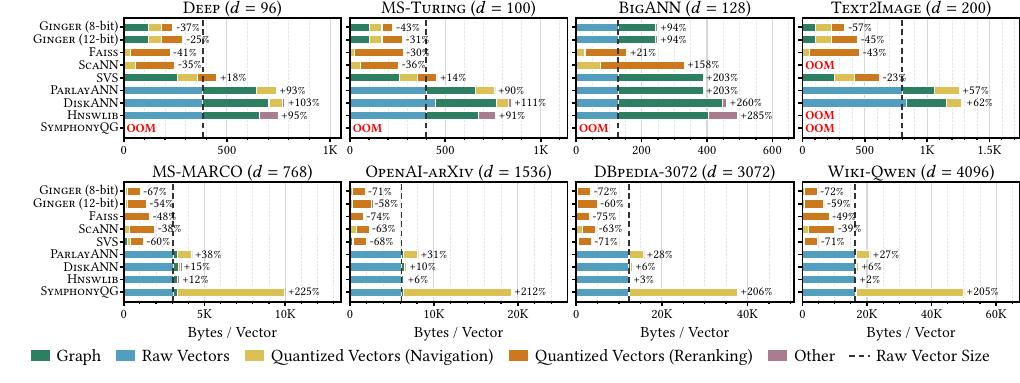}
  \caption{Memory footprint in bytes per vector broken down by index component. The dashed vertical indicates the uncompressed vector size. Percentages show the memory overhead or savings relative to the raw vectors. On high-dimensional datasets, we report \textsc{Ginger} with 8-bit ($4+4$) and 12-bit ($4+8$) NTQ reranking.}
  \label{fig:memory_breakdown}
\end{figure*}

\begin{table*}[t]
  \caption{Index construction times. $^*$Times for \textsc{SVS} include patches to speed up build but still produce the same graph. $^\dagger$Times for \textsc{ScaNN} include patches to significantly speed up build times with k-means on quantized vectors.
  }
  \label{tab:construction_times}
  \centering
  \resizebox{0.95\textwidth}{!}{%
  \begin{tabular}{clcccccccccccc}
    \toprule
    & & \multicolumn{4}{c}{\textsc{Ginger} (\pipnn)} & \textsc{Ginger} (\textsc{Vamana}) & \multirow{2}{*}{\textsc{SVS}$^*$} & \multirow{2}{*}{\textsc{ParlayANN}} & \multirow{2}{*}{\textsc{DiskANN}} & \multirow{2}{*}{\textsc{SymphonyQG}} & \multirow{2}{*}{\textsc{Hnswlib}} & \multirow{2}{*}{\textsc{ScaNN}$^\dagger$} & \multirow{2}{*}{\textsc{Faiss}} \\
    \cmidrule(lr){3-6}
    & Dataset & Quantization \& PCA & Graph Compression & Graph Construction & Total & (No Compression) & & & & & & & \\
    \midrule
    \multirow{6}{*}{\rotatebox[origin=c]{90}{Low dimensional}}
    & \textsc{GloVe} & <1s & 3s & \cellcolor[HTML]{C8E6C9}\textbf{1s} & \cellcolor[HTML]{E9EFCB}4s & \cellcolor[HTML]{F7F4CC}7s & \cellcolor[HTML]{F8CECC}1m 11s & \cellcolor[HTML]{FEF0CC}13s & \cellcolor[HTML]{FAF5CC}8s & \cellcolor[HTML]{FEEECC}14s & \cellcolor[HTML]{FEF1CC}12s & \cellcolor[HTML]{F7F4CC}7s & \cellcolor[HTML]{D5EACA}\textbf{2s} \\
    & \textsc{Deep} & 42s & 16m 3s & \cellcolor[HTML]{C8E6C9}\textbf{17m 44s} & \cellcolor[HTML]{D7EACA}\textbf{34m 30s} & \cellcolor[HTML]{E6EFCB}1h 8m & \cellcolor[HTML]{FCE5CC}9h 31m & \cellcolor[HTML]{F6F3CB}2h 13m & \cellcolor[HTML]{F7F4CC}2h 25m & \cellcolor[HTML]{F8CECC}OOM & \cellcolor[HTML]{FFF5CC}3h 34m & \cellcolor[HTML]{F8CECC}38h 23m & \cellcolor[HTML]{E3EECA}57m 51s \\
    & \textsc{MS-Turing} & 48s & 20m 9s & \cellcolor[HTML]{C8E6C9}\textbf{18m 31s} & \cellcolor[HTML]{D9EBCA}\textbf{39m 28s} & \cellcolor[HTML]{DEECCA}49m 0s & \cellcolor[HTML]{FADBCC}18h 53m & \cellcolor[HTML]{EFF1CB}1h 46m & \cellcolor[HTML]{F0F2CB}1h 51m & \cellcolor[HTML]{F8CECC}OOM & \cellcolor[HTML]{FBF5CC}2h 58m & \cellcolor[HTML]{F8CECC}41h 55m & \cellcolor[HTML]{E4EECB}1h 6m \\
    & \textsc{MS-SPACEV} & -- & 13m 21s & \cellcolor[HTML]{C8E6C9}\textbf{17m 55s} & \cellcolor[HTML]{D4EACA}\textbf{31m 16s} & \cellcolor[HTML]{EEF1CB}1h 37m & \cellcolor[HTML]{FBDECC}16h 4m & \cellcolor[HTML]{F3F3CB}2h 4m & \cellcolor[HTML]{EFF1CB}1h 44m & \cellcolor[HTML]{F8CECC}OOM & \cellcolor[HTML]{FDF5CC}3h 11m & \cellcolor[HTML]{F8CECC}41h 52m & \cellcolor[HTML]{E5EFCB}1h 7m \\
    & \textsc{BigANN} & -- & 16m 12s & \cellcolor[HTML]{C8E6C9}\textbf{18m 18s} & \cellcolor[HTML]{D7EACA}\textbf{34m 30s} & \cellcolor[HTML]{F0F2CB}1h 40m & \cellcolor[HTML]{FDECCC}5h 40m & \cellcolor[HTML]{FDF5CC}2h 47m & \cellcolor[HTML]{F8F4CC}2h 20m & \cellcolor[HTML]{F8CECC}OOM & \cellcolor[HTML]{FFF5CC}3h 14m & \cellcolor[HTML]{F8CECC}31h 17m & \cellcolor[HTML]{E8EFCB}1h 9m \\
    & \textsc{Text2Image} & 1m 30s & 15m 52s & \cellcolor[HTML]{C8E6C9}\textbf{20m 24s} & \cellcolor[HTML]{DAEBCA}\textbf{37m 47s} & \cellcolor[HTML]{E9F0CB}1h 2m & \cellcolor[HTML]{F8CECC}13h 54m & \cellcolor[HTML]{FCE6CC}4h 30m & \cellcolor[HTML]{F7F4CC}1h 41m & \cellcolor[HTML]{F8CECC}OOM & \cellcolor[HTML]{F8CECC}OOM & \cellcolor[HTML]{F8CECC}OOM & \cellcolor[HTML]{F3F3CB}1h 28m \\
    \midrule
    \multirow{8}{*}{\rotatebox[origin=c]{90}{High dimensional}}
    & \textsc{Wiki-Cohere} & 17s & 28s & \cellcolor[HTML]{C8E6C9}\textbf{49s} & \cellcolor[HTML]{DFEDCA}\textbf{1m 34s} & \cellcolor[HTML]{FEF0CC}5m 9s & \cellcolor[HTML]{F8CECC}19m 56s & \cellcolor[HTML]{FCE4CC}8m 16s & \cellcolor[HTML]{FBE1CC}9m 20s & \cellcolor[HTML]{FCE6CC}7m 33s & \cellcolor[HTML]{F8D0CC}18m 37s & \cellcolor[HTML]{F6F3CC}3m 6s & \cellcolor[HTML]{FEF6CC}3m 52s \\
    & \textsc{Wiki-All} & 43s & 1m 13s & \cellcolor[HTML]{C8E6C9}\textbf{2m 0s} & \cellcolor[HTML]{E1EDCA}\textbf{3m 56s} & \cellcolor[HTML]{FEF1CC}10m 45s & \cellcolor[HTML]{F8CECC}40m 25s & \cellcolor[HTML]{FDEBCC}13m 38s & \cellcolor[HTML]{FBE0CC}20m 41s & \cellcolor[HTML]{F8CECC}40m 52s & \cellcolor[HTML]{F8CFCC}39m 59s & \cellcolor[HTML]{FFF3CC}10m 0s & \cellcolor[HTML]{FFF6CC}9m 7s \\
    & \textsc{MS-MARCO} & 50s & 2m 5s & \cellcolor[HTML]{C8E6C9}\textbf{2m 34s} & \cellcolor[HTML]{E1EDCA}\textbf{5m 29s} & \cellcolor[HTML]{FEEECC}18m 53s & \cellcolor[HTML]{F8CECC}1h 10m & \cellcolor[HTML]{FCE2CC}30m 14s & \cellcolor[HTML]{FBE0CC}33m 28s & \cellcolor[HTML]{FAD7CC}47m 17s & \cellcolor[HTML]{F8D1CC}1h 3m & \cellcolor[HTML]{FFF6CC}13m 35s & \cellcolor[HTML]{F5F3CB}9m 47s \\
    & \textsc{DBpedia-1536} & 2s & 2s & \cellcolor[HTML]{C8E6C9}\textbf{2s} & \cellcolor[HTML]{EFF1CB}\textbf{6s} & \cellcolor[HTML]{FEF6CC}9s & \cellcolor[HTML]{FAD8CC}31s & \cellcolor[HTML]{FAF5CC}8s & \cellcolor[HTML]{FBDECC}24s & \cellcolor[HTML]{FBDECC}24s & \cellcolor[HTML]{F9D4CC}36s & \cellcolor[HTML]{FCE4CC}19s & \cellcolor[HTML]{F8CECC}46s \\
    & \textsc{OpenAI-arXiv} & 4s & 3s & \cellcolor[HTML]{C8E6C9}\textbf{4s} & \cellcolor[HTML]{ECF0CB}\textbf{11s} & \cellcolor[HTML]{FEF2CC}22s & \cellcolor[HTML]{F9D3CC}1m 8s & \cellcolor[HTML]{FEF1CC}23s & \cellcolor[HTML]{F9D7CC}1m 0s & \cellcolor[HTML]{FADBCC}50s & \cellcolor[HTML]{F8CECC}1m 22s & \cellcolor[HTML]{FAD9CC}55s & \cellcolor[HTML]{F9D2CC}1m 10s \\
    & \textsc{DataComp-DR} & 15s & 10s & \cellcolor[HTML]{C8E6C9}\textbf{27s} & \cellcolor[HTML]{DEEDCA}\textbf{52s} & \cellcolor[HTML]{FFF5CC}2m 19s & \cellcolor[HTML]{FBDECC}5m 47s & \cellcolor[HTML]{F8CECC}10m 53s & \cellcolor[HTML]{FADBCC}6m 32s & \cellcolor[HTML]{F8D0CC}9m 57s & \cellcolor[HTML]{F9D3CC}8m 49s & \cellcolor[HTML]{FCE5CC}4m 20s & \cellcolor[HTML]{FCE6CC}4m 13s \\
    & \textsc{DBpedia-3072} & 11s & 2s & \cellcolor[HTML]{C8E6C9}\textbf{5s} & \cellcolor[HTML]{FCF5CC}18s & \cellcolor[HTML]{FCE5CC}33s & \cellcolor[HTML]{F8CECC}1m 8s & \cellcolor[HTML]{E8EFCB}\textbf{11s} & \cellcolor[HTML]{FAD9CC}48s & \cellcolor[HTML]{FAD9CC}48s & \cellcolor[HTML]{F8CFCC}1m 6s & \cellcolor[HTML]{FAD7CC}51s & \cellcolor[HTML]{FAD8CC}50s \\
    & \textsc{Wiki-Qwen} & 22s & 1s & \cellcolor[HTML]{C8E6C9}5s & \cellcolor[HTML]{FEEECC}29s & \cellcolor[HTML]{FDE9CC}35s & \cellcolor[HTML]{F9D6CC}1m 10s & \cellcolor[HTML]{C8E6C9}\textbf{5s} & \cellcolor[HTML]{FDEDCC}30s & \cellcolor[HTML]{FDE8CC}36s & \cellcolor[HTML]{FCE6CC}39s & \cellcolor[HTML]{F9D4CC}1m 16s & \cellcolor[HTML]{F8CECC}1m 34s \\
    \bottomrule
  \end{tabular}%
  }
\end{table*}

\subsection{Comparison with other Libraries}

We evaluate the libraries along three dimensions: query throughput, memory footprint, and indexing time. 
These three properties jointly comprise the bedrock of index performance. We use geometric means for comparing across datasets. 
In general: \sys{} produces \textit{smaller indexes} which support \textit{higher query throughput} with \textit{less indexing time} as compared with all other libraries.

\myparagraph{Query Throughput}
Figure~\ref{fig:recall_qps} shows throughput (queries per second) with respect to recall.
\sys{} leads on 13 of 14 datasets across the $[90\%, 100\%]$ recall range and in area under the QPS-recall curve (AUC).
At $95\%$ recall, its mean speedup over the fastest competitor on each dataset is $1.8\times$ overall and $2.0\times$ on high-dimensional datasets, with the largest dataset-specific gain of $2.7\times$ on \textsc{Wiki-Cohere}.
The strongest competitors on high-dimensional datasets are \textsc{SymphonyQG} and \textsc{SVS}.
The only exception is \textsc{BigANN}, whose byte coordinates we do not encode with TurboQuant; there, \textsc{SVS} is $4\%$ faster at both $90\%$ and $95\%$ recall.
Because different competitors are second-best on different datasets, the advantage over any individual library across completed datasets is larger.
In AUC, \sys{} achieves a geometric mean gain of $2.35\times$ over \textsc{ParlayANN}, $2.38\times$ over \textsc{SVS}, $2.42\times$ over \textsc{SymphonyQG}, and $2.79\times$ over \textsc{DiskANN}.

The IVF indexes \textsc{Faiss} and \textsc{ScaNN} trail the graph methods across all datasets, with \sys{} achieving a geometric mean AUC gain of $10.3\times$ over \textsc{ScaNN} and $11.7\times$ over \textsc{Faiss}.

\myparagraph{Memory Footprint}
The memory comparison exposes two different sources of savings: graph compression on low-dimensional datasets and vector quantization on high-dimensional datasets.
In Figure~\ref{fig:memory_breakdown}, we compare the memory footprints. 
We split each index into five parts.
\emph{Graph} is the adjacency structure, which only \textsc{Ginger} compresses.
\emph{Raw Vectors} are the uncompressed vectors, for indexes that keep them.
\emph{Quantized Vectors (Navigation)} are the codes scored during graph traversal or IVF scanning, including JL sketches and, for IVF indexes, the inverted lists and the coarse quantizer.
\emph{Quantized Vectors (Reranking)} are the higher-precision codes used to rerank candidates instead of raw vectors.
\emph{Other} covers ID remapping tables and locks.

\begin{figure*}[t]
  \centering
  \includegraphics[width=0.9\textwidth, trim=0 8pt 0 0, clip]{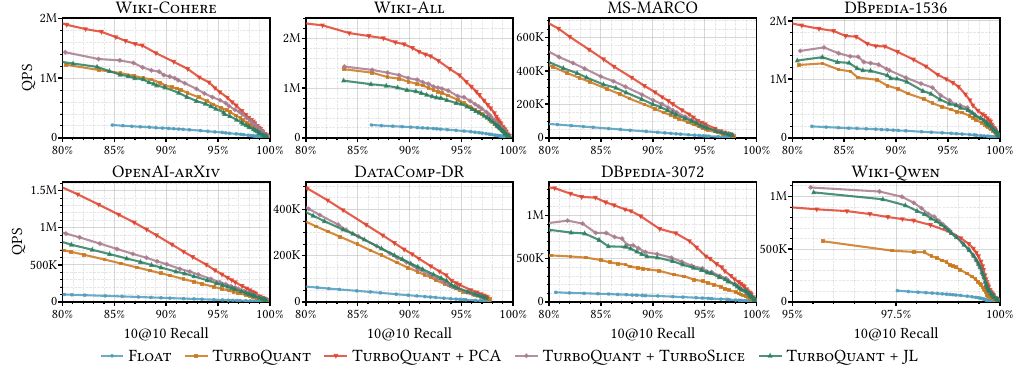}
   \caption{Ablation comparing \textsc{Ginger} search variants. TurboSlice uses $d_{\mathrm{ts}} = 256$, JL uses $d_{\mathrm{jl}} = 1024$.}
  \label{fig:recall_qps_candidate_prefilter}
\end{figure*}

On the four low-dimensional datasets in Figure~\ref{fig:memory_breakdown}, \textsc{Ginger} has the smallest index of all graph-based methods.
It is on average $1.54\times$ (up to $1.65\times$) smaller than \textsc{SVS}, the most compact competitor.
Most of this saving comes from graph compression.
The graph of \textsc{Ginger} takes on average $111$ bytes per vector, compared with $260$ bytes for the flat adjacency arrays of \textsc{SVS} and \textsc{ParlayANN}.
Only the IVF indexes \textsc{Faiss} and \textsc{ScaNN} are competitive with regards to memory footprint on low-dimensional datasets.

On high-dimensional datasets the embeddings themselves dominate the memory footprint.
\textsc{Ginger} needs on average $43\%$ of the raw vector size across all eight high-dimensional datasets due to NTQ with 12 bits.
\textsc{ParlayANN}, \textsc{DiskANN}, and \textsc{Hnswlib} keep the raw float vectors and are $2.7\times$ larger.
\textsc{SymphonyQG} is $7.4\times$ larger because of its per-edge codes.
\textsc{SVS}, \textsc{Faiss}, and \textsc{ScaNN} are smaller on some datasets, by up to $1.6\times$ (\textsc{Faiss} on \textsc{OpenAI-arXiv}), mainly because they rerank with 8-bit instead of 12-bit codes.
\textsc{SVS} uses $19.8\%$ less memory in the geometric mean across all eight high-dimensional datasets.
However, \textsc{SVS} is $2.9\times$ slower in the geometric mean at $95\%$ recall.
It also reaches lower peak recall than 12-bit \textsc{Ginger} on every dataset.
With 8-bit NTQ, \textsc{Ginger} uses less memory than \textsc{SVS} on all datasets, still achieves $2.6\times$ higher mean QPS at $95\%$ recall and higher peak recall on seven of the eight datasets.

\myparagraph{Indexing Time}
These memory savings require additional pre- and post-processing, so we compare complete index construction times, including graph reordering, compression, quantization, and PCA.
Table~\ref{tab:construction_times} breaks down \sys{} (\pipnn) builds into graph construction, graph compression, and quantization plus PCA.
Fast graph construction comes from \pipnn~\cite{rubel2026pipnn,rubel2026turbocharging}; quantization is fast and grouped with PCA to save space.

With \pipnn, \sys{} builds fastest on 11 of 14 datasets, achieving a mean speedup of $1.4\times$ over the fastest competitor (up to $4.9\times$ on \textsc{DataComp-DR}).
Note that these times include graph reordering and compression for Ginger, but not the other methods, trading some of the advantage of PiPNN for lower memory footprint.
On the five billion-scale datasets, complete builds take an average of $35$ minutes (up to $39$ minutes), including $18$ to $20$ minutes for graph construction and $13$ to $20$ minutes for reordering and compression.
\textsc{Faiss}, the fastest competitor, takes on average $69$ minutes (up to $88$ minutes), giving \sys{} a $2.0\times$ mean speedup (up to $2.3\times$).
The remaining libraries require $1.8$ to $42$ hours or run out of memory.
On the three large high-dimensional datasets ($35$M to $101$M vectors), \sys{} builds $2.0\times$ faster in the geometric mean (up to $2.3\times$) than the fastest competitor.

On smaller datasets, all methods finish within two minutes.
\textsc{ParlayANN} builds up to $5.7\times$ faster than \sys{} on the very small datasets \textsc{DBpedia-3072} and \textsc{Wiki-Qwen}, where PCA dominates our build time, while \textsc{Faiss} is $2.1\times$ faster on \textsc{GloVe}.
If we look only at \text{PiPNN} graph construction time, it is fastest.

The \sys{} (\textsc{Vamana}) column reports our Vamana implementation without graph compression for reference.
Even with \textsc{Vamana}, \sys{} builds faster than all external \textsc{Vamana} libraries on the 11 datasets with $n > 1\text{M}$.

To be able to include them in the experiments, we patched \textsc{SVS} and \textsc{ScaNN} to accelerate index construction. For \textsc{SVS}, we eliminate lock contention when adding backward edges to the Vamana graph by scaling the number of backedge buffers and clearing them in parallel, giving greater than $2\times$ speedup without changing the resulting graph. For \textsc{ScaNN}, we parallelize several sequential loops and accelerate k-means by running on quantized 8-bit vectors (SFP8), which has been shown to not compromise k-means quality~\cite{kuffo2026superfastkmeans}.

\subsection{Ablation Studies}\label{sec:ablations}
We now analyze the contributions of \sys{}'s main components through ablation studies.
We first examine search optimizations, then evaluate NTQ accuracy independently of graph traversal, and finally study how reordering and encoding affect graph size and query speed.

\begin{figure}[b]
  \centering
  \includegraphics[width=0.95\columnwidth, trim=0 7.2pt 0 0, clip]{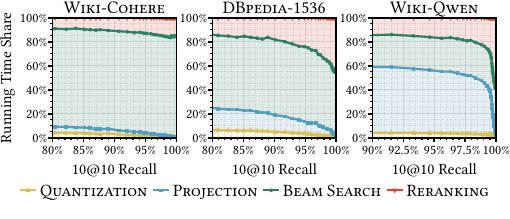}
  \caption{Share of query time spent in each search phase. On higher-dimensional datasets, projection accounts for a growing share of query time due to the larger projection matrix.}
  \label{fig:recall_qps_breakdown}
\end{figure}

\myparagraph{Search Optimizations}
Figure~\ref{fig:recall_qps_candidate_prefilter} breaks down where \textsc{Ginger}'s speedups come from.
At $90\%$ recall, navigating on 4-bit TurboQuant codes instead of float vectors is the largest step, speeding up search by $5.7\times$ in the geometric mean (up to $7.2\times$).
Searching on PCA-projected coordinates speeds up search by another $1.8\times$ in the geometric mean (up to $2.4\times$ on \textsc{DBpedia-3072}, which has the largest reduction in dimension).
The prefilters help to a lesser extent at $90\%$ recall.
TurboSlice speeds up search by $1.26\times$ in the geometric mean (up to $1.5\times$).
JL speeds up search by $1.12\times$ in the geometric mean (up to $1.4\times$ on \textsc{DBpedia-3072}), but incurs a $15\%$ slowdown on \textsc{Wiki-All}.
TurboSlice beats JL on six of the seven datasets.
On \textsc{Wiki-Qwen}, TurboSlice is $1.27\times$ faster than PCA at $97\%$ recall, and the two are on par at $99\%$ recall.
We attribute this to the fact that the projection is expensive on \textsc{Wiki-Qwen}.
PCA and the prefilters both cut memory traffic per candidate, so their gains do not stack.

\begin{table*}[t]
\footnotesize
  \setlength{\aboverulesep}{0pt}
  \setlength{\belowrulesep}{0pt}
  \renewcommand{\arraystretch}{1.0}
  \caption{Graph reordering quality and runtime. We report $\textsf{LogGap} = \frac{1}{|E|}\textstyle\sum \log_2(g_i+1)$ and actual encoded bits per edge under \textsf{AdaptiveGap} for the neighbor lists only, excluding the CSR row-offset array. Cells are shaded per dataset from green (best) to red (worst).}
  \label{tab:reordering_ablation}
  \centering
  \resizebox{0.95\textwidth}{!}{%
  \begin{tabular}{ll cccccc}
    \toprule
    Metric & Reordering Method & \textsc{OpenAI-arXiv} (2.3M) & \textsc{Wiki-Cohere} (35M) & \textsc{Wiki-All} (88M) & \textsc{Text2Image} (100M) & \textsc{Deep} (100M) & \textsc{MS-Turing} (100M) \\
    \midrule
  \multirow{4}{*}{\shortstack[l]{$\textsf{LogGap} \;/\; \textsf{AdaptiveGap}$\\(bits/edge)}}
    & \textsf{Unreordered} & \cellcolor[HTML]{F8CECC}14.53 / 19.34 & \cellcolor[HTML]{F8CECC}16.88 / 22.05 & \cellcolor[HTML]{F8CECC}16.78 / 22.43 & \cellcolor[HTML]{F8CECC}19.95 / 24.55 & \cellcolor[HTML]{F8CECC}20.14 / 24.85 & \cellcolor[HTML]{F8CECC}20.22 / 24.93 \\
    & \textsf{RGC} & \cellcolor[HTML]{FFF6CC}10.10 / 15.66 & \cellcolor[HTML]{FFF6CC}12.84 / 18.38 & \cellcolor[HTML]{FCE2CC}13.89 / 19.69 & \cellcolor[HTML]{FCE2CC}9.71 / 15.22 & \cellcolor[HTML]{FCE2CC}15.04 / 20.55 & \cellcolor[HTML]{FCE2CC}15.19 / 20.63 \\
    & \textsf{RGB}~\cite{dhulipala2016compressing} & \cellcolor[HTML]{C8E6C9}8.80 / 14.17 & \cellcolor[HTML]{E4EECA}11.08 / 16.54 & \cellcolor[HTML]{FFF6CC}12.01 / 17.82 & \cellcolor[HTML]{FFF6CC}8.65 / 13.99 & \cellcolor[HTML]{C8E6C9}11.99 / 17.51 & \cellcolor[HTML]{FFF6CC}13.18 / 18.79 \\
    & \textsf{RGP} (Ours) & \cellcolor[HTML]{C8E6C9}\textbf{8.57 / 13.97} & \cellcolor[HTML]{C8E6C9}\textbf{10.59 / 16.02} & \cellcolor[HTML]{C8E6C9}\textbf{11.02 / 16.86} & \cellcolor[HTML]{C8E6C9}\textbf{7.81 / 13.08} & \cellcolor[HTML]{C8E6C9}\textbf{11.74 / 17.27} & \cellcolor[HTML]{C8E6C9}\textbf{12.03 / 17.63} \\
    \midrule
    \multirow{3}{*}{Reorder Time (s)}
    & \textsf{RGC} & \cellcolor[HTML]{C8E6C9}\textbf{2.4s} & \cellcolor[HTML]{FCE2CC}83.2s & \cellcolor[HTML]{F8CECC}1573.4s & \cellcolor[HTML]{FCE2CC}248.3s & \cellcolor[HTML]{FFF6CC}210.7s & \cellcolor[HTML]{FFF6CC}189.5s \\
    & \textsf{RGB}~\cite{dhulipala2016compressing} & \cellcolor[HTML]{E4EECA}3.5s & \cellcolor[HTML]{FCE2CC}103.8s & \cellcolor[HTML]{FCE2CC}213.1s & \cellcolor[HTML]{F8CECC}521.5s & \cellcolor[HTML]{FCE2CC}458.6s & \cellcolor[HTML]{FCE2CC}474.4s \\
    & \textsf{RGP} (Ours) & \cellcolor[HTML]{E4EECA}3.2s & \cellcolor[HTML]{C8E6C9}\textbf{20.0s} & \cellcolor[HTML]{C8E6C9}\textbf{59.8s} & \cellcolor[HTML]{C8E6C9}\textbf{79.6s} & \cellcolor[HTML]{C8E6C9}\textbf{91.4s} & \cellcolor[HTML]{C8E6C9}\textbf{113.3s} \\
    \bottomrule
  \end{tabular}%
  }
\end{table*}

\begin{table*}[t]
\footnotesize
  \setlength{\aboverulesep}{0pt}
  \setlength{\belowrulesep}{0pt}
  \renewcommand{\arraystretch}{1.0}
  \caption{QPS at $\ge 0.95$ recall / graph memory (bits per edge, including CSR offsets). Cells are shaded per dataset column by QPS from green (highest) to red (lowest). Note that bits per edge is higher than in Table~\ref{tab:reordering_ablation} due to including CSR offsets.}
  \label{tab:codec_ablation}
  \centering
  \resizebox{0.95\textwidth}{!}{%
  \begin{tabular}{l cccccc}
    \toprule
    Graph Format \& Reordering & \textsc{OpenAI-arXiv} (2.3M) & \textsc{Wiki-Cohere} (35M) & \textsc{Wiki-All} (88M) & \textsc{Text2Image} (100M) & \textsc{Deep} (100M) & \textsc{MS-Turing} (100M) \\
    \midrule
    \textsf{AdaptiveGap + Reordered (CSR)} & \cellcolor[HTML]{C8E6C9}519.5k / \textbf{14.5b} & \cellcolor[HTML]{E4EECA}721.4k / \textbf{16.5b} & \cellcolor[HTML]{E4EECA}1149.9k / \textbf{17.6b} & \cellcolor[HTML]{C8E6C9}118.5k / \textbf{13.4b} & \cellcolor[HTML]{E4EECA}775.9k / \textbf{17.7b} & \cellcolor[HTML]{E4EECA}501.9k / \textbf{18.1b} \\
    \textsf{StreamVByte + Reordered (CSR)} & \cellcolor[HTML]{C8E6C9}\textbf{527.9k} / 15.7b & \cellcolor[HTML]{E4EECA}723.3k / 17.4b & \cellcolor[HTML]{E4EECA}1160.6k / 18.6b & \cellcolor[HTML]{C8E6C9}117.9k / 14.3b & \cellcolor[HTML]{E4EECA}782.2k / 18.5b & \cellcolor[HTML]{E4EECA}511.2k / 18.7b \\
    \textsf{PrefixVarint + Reordered (CSR)} & \cellcolor[HTML]{FFF6CC}499.6k / 14.7b & \cellcolor[HTML]{FFF6CC}693.9k / 17.0b & \cellcolor[HTML]{FFF6CC}1098.1k / 18.0b & \cellcolor[HTML]{FFF6CC}111.8k / 13.6b & \cellcolor[HTML]{FCE2CC}688.4k / 18.1b & \cellcolor[HTML]{FCE2CC}471.8k / 18.6b \\
    \midrule
    \textsf{Uncompressed + Reordered (FlatGraph)} & \cellcolor[HTML]{C8E6C9}523.9k / 65.2b & \cellcolor[HTML]{C8E6C9}740.9k / 53.9b & \cellcolor[HTML]{C8E6C9}1195.5k / 89.3b & \cellcolor[HTML]{C8E6C9}119.6k / 39.2b & \cellcolor[HTML]{C8E6C9}\textbf{807.4k} / 48.1b & \cellcolor[HTML]{C8E6C9}\textbf{527.0k} / 49.6b \\
    \textsf{Uncompressed (FlatGraph)} & \cellcolor[HTML]{C8E6C9}526.9k / 65.2b & \cellcolor[HTML]{C8E6C9}\textbf{747.0k} / 53.9b & \cellcolor[HTML]{C8E6C9}\textbf{1203.2k} / 89.3b & \cellcolor[HTML]{C8E6C9}119.2k / 39.2b & \cellcolor[HTML]{FFF6CC}745.5k / 48.1b & \cellcolor[HTML]{C8E6C9}525.6k / 49.6b \\
    \midrule
    \textsf{Uncompressed + Reordered (CSRGraph)} & \cellcolor[HTML]{C8E6C9}520.9k / 34.0b & \cellcolor[HTML]{E4EECA}726.2k / 33.7b & \cellcolor[HTML]{C8E6C9}1180.0k / 34.7b & \cellcolor[HTML]{C8E6C9}119.7k / 33.2b & \cellcolor[HTML]{FFF6CC}752.1k / 33.5b & \cellcolor[HTML]{E4EECA}515.1k / 33.5b \\
    \textsf{Uncompressed (CSRGraph)} & \cellcolor[HTML]{C8E6C9}523.3k / 34.0b & \cellcolor[HTML]{E4EECA}731.9k / 33.7b & \cellcolor[HTML]{E4EECA}1176.7k / 34.7b & \cellcolor[HTML]{C8E6C9}\textbf{120.1k} / 33.2b & \cellcolor[HTML]{E4EECA}777.3k / 33.5b & \cellcolor[HTML]{E4EECA}515.0k / 33.5b \\
    \bottomrule
  \end{tabular}%
  }
\end{table*}

\begin{figure}[b]
  \centering
  \includegraphics[width=0.95\columnwidth, trim=0 7.3pt 0 0, clip]{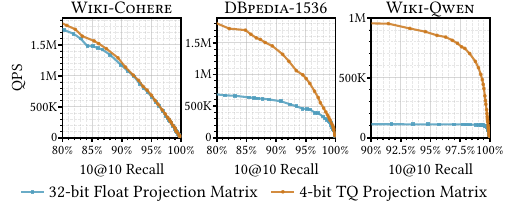}
  \caption{Quantizing the PCA matrix improves QPS on very high dimensions.}
  \label{fig:recall_qps_pca_matrix_repr_3ds}
\end{figure}

Next, we investigate how much time each phase of the search procedure takes.
Figure~\ref{fig:recall_qps_breakdown} shows how the runtime balance shifts as dimensionality increases from $768$ (\textsc{Wiki-Cohere}) to $1{,}536$ (\textsc{DBpedia-1536}) and $4{,}096$ (\textsc{Wiki-Qwen}).
Beam search dominates on the first two datasets, while projection accounts for roughly half the runtime on \textsc{Wiki-Qwen} at $95\%$ recall.
This reflects the growing cost of multiplying each query by a projection matrix whose size increases with both the input and projected dimensions.
On larger dimensionalities, we can significantly speed up projection without sacrificing recall by quantizing the matrix entries.
Figure~\ref{fig:recall_qps_pca_matrix_repr_3ds} shows little benefit from quantizing \textsc{Wiki-Cohere}'s projection matrix, whose float representation occupies only $0.73$\,MiB and can fit in L2 cache.
The larger float matrices for \textsc{DBpedia-1536} and \textsc{Wiki-Qwen} occupy $2.16$\,MiB and $13.75$\,MiB, respectively.
Quantization reduces their code storage to $0.27$\,MiB and $1.72$\,MiB, consistent with improved cache residency and the substantial throughput gains on these datasets.
The benefit narrows near full recall as the fixed query-projection cost is amortized over more expensive search and reranking.

\begin{figure}[b]
  \centering
  \includegraphics[width=\columnwidth, trim=0 4pt 0 0, clip]{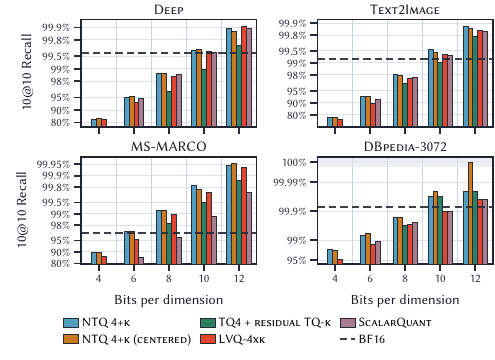}
  \caption{Exhaustive search $10@10$ recall on four datasets (subsetted to 5M datapoints) across bitwidths. Showing \texttt{bf16} as a horizontal dashed line.
  }
  \label{fig:quant_recall_10at10}
\end{figure}

\myparagraph{Quantization Ablations}\phantomsection\label{sec:quant-ablations}
In Figure~\ref{fig:quant_recall_10at10} we plot the recall of NTQ at different bitwidths, showing that 12 bits achieves near float accuracy. 
We measure the accuracy of the quantizer separate from the quality of an approximate search index by evaluating recall over a brute-force scoring of the dataset.
We use subsets of 5 million datapoints per dataset so as to make this computationally tractable.
Moreover, these results show that a simpler residual scheme, which quantizes $r = y - \tilde{y}$ with a second TurboQuant, plain uniform scalar quantization, and \texttt{bf16} are all insufficient.
We also compare with a reimplementation of Intel's proprietary LVQ codec~\cite{aguerrebere2023similarity}.
Like NTQ, two-level LVQ (LVQ-4x$k$) pairs a 4-bit navigation code with $k$ additional bits per coordinate for reranking.
For LVQ-4x8, the recall of our reimplementation agrees with Intel's SVS binaries to four decimal places.
LVQ centers each vector by the global dataset mean and quantizes it uniformly to 4 bits with per-vector minimum and maximum bounds.
For reranking, the residual of the centered vector with respect to its 4-bit code is requantized with uniform scalar quantization.
At 4 bits, LVQ is less accurate than NTQ on the three high-dimensional datasets, while both approach float accuracy at 12 bits.
The use of the global mean is crucial for LVQ's recall which makes the codec less flexible in dynamic settings.
Adding centering to NTQ can optionally boost recall on some datasets, such as \textsc{DBpedia-3072}.

Extended RaBitQ~\cite{gao2025extendedrabitq} is another method that splits its codes, into a 1-bit sign and a $k$-bit magnitude.
Its only coarse level is thus the 1-bit sign code.
Truncating the magnitude bits does not yield a well-tuned coarse code either, because all bits share one uniform grid whose per-vector scale is fit to the full code.

\myparagraph{Graph Compression and Reordering}\phantomsection\label{sec:compression-ablations}
Lastly, we evaluate the impact of graph reordering on compression and query speed.
Table~\ref{tab:reordering_ablation} compares \textsf{RGP} (Section~\ref{sec:compress:reordering}) against the original vertex order (\textsf{Unreordered}), and two baselines from the literature: bottom-up agglomerative clustering on normalized edge weights (\textsf{RGC}~\cite{blandford2003compact}), and top-down Recursive Graph Bisection (\textsf{RGB}~\cite{dhulipala2016compressing}).
\textsf{RGP} achieves the smallest gaps on all six datasets and reduces the bits per edge under \textsf{AdaptiveGap} by $31.5\%$ in the geometric mean (up to $46.7\%$) over the original order.
On the larger graphs ($35$M to $100$M vertices), \textsf{RGP} is also $4.2\times$ faster in the geometric mean (up to $26.3\times$) than \textsf{RGC} and $4.8\times$ faster in the geometric mean (up to $6.6\times$) than \textsf{RGB}.

The graph-memory measurements in Table~\ref{tab:codec_ablation} include the CSR row-offset array, whereas Table~\ref{tab:reordering_ablation} reports encoded neighbor-list storage alone.
Table~\ref{tab:codec_ablation} compares query throughput at $95\%$ recall and graph memory for three compression codecs on \textsf{RGP}-reordered graphs (\textsf{AdaptiveGap}, \textsf{StreamVByte}~\cite{lemire2018streamvbyte}, and \textsf{PrefixVarint}~\cite{dean2009challenges}) against uncompressed CSR (\textsf{CSRGraph}) and fixed-degree arrays (\textsf{FlatGraph}).
\textsf{AdaptiveGap} shrinks the graph by $2.1\times$ in the geometric mean (up to $2.5\times$) over \textsf{CSRGraph} and by $3.4\times$ in the geometric mean (up to $5.1\times$) over \textsf{FlatGraph}, compared with $2.0\times$ and $3.2\times$ in the geometric mean for \textsf{StreamVByte}.
\textsf{AdaptiveGap} comes within $1.4\%$ of \textsf{CSRGraph} and $2.8\%$ of \textsf{FlatGraph} in the geometric mean while using $2.1\times$ and $3.4\times$ less graph memory.
Non-vectorized decoding makes \textsf{PrefixVarint} $5.9\%$ slower in the geometric mean (up to $11.3\%$) than \textsf{AdaptiveGap}.
\textsf{AdaptiveGap} matches the throughput of \textsf{StreamVByte} while saving $0.9$ bits per edge in the geometric mean (up to $1.2$ bits), or $5.4\%$ of graph memory (up to $7.6\%$).

\section{Discussion \& Conclusion}
\label{sec:conc}

\sys{} shows that graph-based vector search still has plenty of room to grow.
It breaks new ground in query throughput, memory footprint, and indexing time for ANNS.
The key is designing every step of the search around reducing memory traffic.
Nested TurboQuant lets a single code serve both fast 4-bit navigation and reranking with near-float accuracy.
Graph compression shrinks the graph without slowing down queries.
PiPNN completes the picture with fast index construction.

\myparagraph{Looking Ahead}
While this paper focuses on single-vector dense embeddings, retrieval applications at Google increasingly go beyond this setting.
We therefore designed \sys{} to be flexible for more general retrieval settings.
Graph construction and beam search operate only through a user-provided similarity function, abstracting away the specific data representation.
This makes it easy to search under novel similarity functions like Chamfer on multi-vectors and black-box similarity functions such as learned relevance scorers.
Neural scorer inference is more expensive than computing a dot product, increasing the appeal of the few distance evaluations that graph search needs.
We are thrilled to share \sys{} with the research community and look forward to seeing what others build with it.

\begin{acks}
 We thank Richard Wen, Kishen Gowda, Majid Daliri, Rasto Lenhardt, Jirka Isa, Jan Wassenberg, Peter Sanders, John Owens and Daniel Seemaier for helpful discussions.
\end{acks}


\bibliographystyle{ACM-Reference-Format}
\bibliography{references}

\appendix

\clearpage

\section{Parameter Configuration}
\label{sec:param-config}

For each library, we tune index structure, quantization method, and reranking formats for the highest throughput across the recall range.
Every configuration we report is built and evaluated on the full dataset.
We sweep query search parameters to extract the Pareto frontier of throughput versus recall.
Table~\ref{tab:query_configs} summarizes the formats we use for each library, and Table~\ref{tab:construction_params} lists the selected parameters for all 14 datasets.
Below, we first summarize the settings for graph indexes.
For the IVF indexes (\textsc{ScaNN} and \textsc{Faiss}), point clustering, routing structure on top of the clusters, quantization, and reranking interact closely, making them hard to tune, so we describe their tuning in detail below.

\begin{table*}[t]
  \caption{Vector representations across libraries. On the integer datasets (\textsc{MS-SPACEV} and \textsc{BigANN}), the graph indexes operate directly on native \textsf{int8}/\textsf{uint8} vectors. $^\ddagger$\textsc{DiskANN} always builds its graph with 4-bit spherical codes. It navigates with 4-bit codes in low dimensions and with 1-bit codes in high dimensions, except for 2-bit codes on \textsc{DataComp-DR}. $^\S$\textsc{SVS} reranks with \textsc{LVQ4x8} in low dimensions and with 8-bit LVQ in high dimensions.}
  \label{tab:query_configs}
  \centering
  \footnotesize
  \begin{tabular}{lccc}
    \toprule
    Library & High-Dim Transform & Navigation / Search & Reranking \\
    \midrule
    \textsc{Ginger} & PCA & 4-bit NTQ & (4+8)-bit NTQ \\
    \textsc{SVS} & PCA (LeanVec) & 4-bit LVQ & LVQ4x8 / 8-bit LVQ$^\S$ \\
    \textsc{ParlayANN} & JL filter & SQ8 & float \\
    \textsc{DiskANN} & -- & 4-bit / 1-bit spherical$^\ddagger$ & float \\
    \textsc{Hnswlib} & -- & float & -- \\
    \textsc{SymphonyQG} & -- & RaBitQ & float \\
    \textsc{ScaNN} & PCA & 2-bit PQ & float / \textsf{bf16} / \textsf{int8} \\
    \textsc{Faiss} & Projective OPQ & 2-bit PQ & float / \textsf{bf16} / \textsf{int8} \\
    \bottomrule
  \end{tabular}
\end{table*}

\begin{table*}[t]
  \centering
  \caption{Dataset-specific index and quantization parameters: robust pruning thresholds $\alpha_{\textsc{PiPNN}}$ and $\alpha_{\textsc{Vamana}}$, reduced PCA dimensionality, \textsc{ScaNN} leaf, intermediate, and top-level cluster counts ($C_1, C_2, C_3$) and reranking formats, and \textsc{Faiss} factory stages (\textsf{OPQ}, \textsf{IVF}, \textsf{PQ}, and \textsf{Refine}). $^\dagger$\textsc{ScaNN} ran out of memory during index construction on \textsc{Text2Image}.}
  \label{tab:construction_params}
  \footnotesize
  \setlength{\tabcolsep}{2.5pt}
  \resizebox{\textwidth}{!}{%
  \begin{tabular}{cl cc ccc rrr c llll}
    \toprule
    & \multirow{2}{*}{Dataset} & \multirow{2}{*}{$\alpha_{\textsc{PiPNN}}$} & \multirow{2}{*}{$\alpha_{\textsc{Vamana}}$} & \multicolumn{3}{c}{PCA} & \multicolumn{4}{c}{\textsc{ScaNN}} & \multicolumn{4}{c}{\textsc{Faiss} Factory Stages} \\
    \cmidrule(lr){5-7} \cmidrule(lr){8-11} \cmidrule(lr){12-15}
    & & & & \textsc{Ginger} & \textsc{SVS} & \textsc{ScaNN} & \multicolumn{1}{c}{$C_1$} & \multicolumn{1}{c}{$C_2$} & \multicolumn{1}{c}{$C_3$} & Rerank & OPQ & IVF & PQ & Refine \\
    \midrule
    \multirow{6}{*}{\rotatebox[origin=c]{90}{Low dimensional}}
    & \textsc{GloVe} & 0.87 & 0.95 & -- & -- & -- & 2{,}000 & \multicolumn{1}{c}{--} & \multicolumn{1}{c}{--} & \textsf{float} & -- & \textsf{IVF1024} & \textsf{PQ50x4fsr} & \textsf{RFlat} \\
    & \textsc{Deep} & 1.15 & 1.10 & -- & -- & -- & 2{,}398{,}082 & 17{,}916 & 134 & \textsf{bf16} & \textsf{OPQ48\_96} & \textsf{IVF1048576\_HNSW32} & \textsf{PQ48x4fsr} & \textsf{Refine(SQbf16)} \\
    & \textsc{MS-Turing} & 1.08 & 1.10 & -- & -- & -- & 2{,}500{,}000 & 18{,}420 & 136 & \textsf{bf16} & \textsf{OPQ50\_100} & \textsf{IVF1048576\_HNSW32} & \textsf{PQ50x4fsr} & \textsf{Refine(Pad128,SQbf16)} \\
    & \textsc{MS-SPACEV} & 1.13 & 1.10 & -- & -- & -- & 2{,}500{,}000 & 18{,}420 & 136 & \textsf{int8} & \textsf{OPQ50\_100} & \textsf{IVF1048576\_HNSW32} & \textsf{PQ50x4fsr} & \textsf{Refine(Pad128,SQ8\_direct\_signed)} \\
    & \textsc{BigANN} & 1.15 & 1.10 & -- & -- & -- & 3{,}205{,}129 & 21{,}739 & 147 & \textsf{bf16} & \textsf{OPQ64\_128} & \textsf{IVF1048576\_HNSW32} & \textsf{PQ64x4fsr} & \textsf{Refine(SQ8\_direct)} \\
    & \textsc{Text2Image}$^\dagger$ & 0.95 & 1.00 & -- & -- & -- & 5{,}000{,}000 & \multicolumn{1}{c}{--} & \multicolumn{1}{c}{--} & -- & \textsf{OPQ100\_200} & \textsf{IVF1048576\_HNSW32} & \textsf{PQ100x4fsr} & \textsf{Refine(SQbf16)} \\
    \midrule
    \multirow{8}{*}{\rotatebox[origin=c]{90}{High dimensional}}
    & \textsc{Wiki-Cohere} & 0.98 & 0.98 & 248 & 256 & 192 & 35{,}000 & 350 & \multicolumn{1}{c}{--} & \textsf{bf16} & \textsf{OPQ96\_192} & \textsf{IVF65536\_HNSW32} & \textsf{PQ96x4fs} & \textsf{Refine(SQbf16)} \\
    & \textsc{Wiki-All} & 0.93 & 0.95 & 248 & 256 & 192 & 87{,}556 & 1{,}369 & \multicolumn{1}{c}{--} & \textsf{bf16} & \textsf{OPQ96\_192} & \textsf{IVF131072\_HNSW32} & \textsf{PQ96x4fs} & \textsf{Refine(SQbf16)} \\
    & \textsc{MS-MARCO} & 0.96 & 0.98 & 248 & 256 & 256 & 101{,}071 & 1{,}580 & \multicolumn{1}{c}{--} & \textsf{bf16} & \textsf{OPQ128\_256} & \textsf{IVF16384\_HNSW32} & \textsf{PQ128x4fs} & \textsf{Refine(SQbf16)} \\
    & \textsc{DBpedia-1536} & 1.05 & 1.00 & 368 & 256 & 256 & 2{,}048 & \multicolumn{1}{c}{--} & \multicolumn{1}{c}{--} & \textsf{int8} & \textsf{OPQ96\_192} & \textsf{IVF2048} & \textsf{PQ96x4fs} & \textsf{Refine(SQ8)} \\
    & \textsc{OpenAI-arXiv} & 1.10 & 1.05 & 368 & 256 & 446 & 10{,}000 & 300 & \multicolumn{1}{c}{--} & \textsf{int8} & \textsf{OPQ128\_256} & \textsf{IVF4096} & \textsf{PQ128x4fs} & \textsf{Refine(SQ8)} \\
    & \textsc{DataComp-DR} & 1.15 & 1.15 & 624 & 640 & 512 & 12{,}805 & 201 & \multicolumn{1}{c}{--} & \textsf{bf16} & \textsf{OPQ256\_512} & \textsf{IVF98304\_HNSW32} & \textsf{PQ256x4fs} & \textsf{Refine(SQbf16)} \\
    & \textsc{DBpedia-3072} & 1.05 & 1.00 & 496 & 256 & 384 & 2{,}048 & \multicolumn{1}{c}{--} & \multicolumn{1}{c}{--} & \textsf{int8} & \textsf{OPQ96\_192} & \textsf{IVF2048} & \textsf{PQ96x4fs} & \textsf{Refine(SQ8)} \\
    & \textsc{Wiki-Qwen} & 0.95 & 1.00 & 880 & 768 & 384 & 1{,}024 & \multicolumn{1}{c}{--} & \multicolumn{1}{c}{--} & \textsf{bf16} & \textsf{OPQ192\_384} & \textsf{IVF2048} & \textsf{PQ192x4fs} & \textsf{Refine(SQbf16)} \\
    \bottomrule
  \end{tabular}%
  }
\end{table*}

\myparagraph{Graph Indexes}
All graph indexes use maximum degree $R = 64$ and construction beam width $L = 128$, except on \textsc{GloVe}, where they use $R = 100$ and $L = 250$.
We tune the pruning threshold $\alpha$ per dataset, with $\alpha_{\textsc{PiPNN}}$ for \textsc{PiPNN} graphs and a single $\alpha_{\textsc{Vamana}}$ shared by all \textsc{Vamana}-based indexes.
Table~\ref{tab:query_configs} lists the quantization methods and formats of each graph index.
\textsc{SymphonyQG} runs out of memory during index construction on the billion-scale datasets, so we evaluate it only on \textsc{GloVe} and the eight high-dimensional datasets.

On high-dimensional datasets ($d \geq 768$), we initialize the PCA target dimension for \sys{} to the smallest dimension that retains roughly $90\%$ of dataset variance, rounded so that the 4-bit NTQ vector and its bundled metadata fill whole cache lines.
Starting from there, we reduce the dimension in increments of whole cache lines to find the optimal throughput--recall trade-off against unprojected vectors.
For \textsc{SVS}, we use \textsc{LeanVec4x8} and follow the same tuning procedure, aligning dimensions directly to whole cache lines as it does not embed metadata into the codes.
Table~\ref{tab:construction_params} lists the chosen dimensions.

\myparagraph{ScaNN}
\textsc{ScaNN} partitions the dataset with a $k$-means tree.
We evaluate up to three levels, with $C_1$ clusters on the bottom, $C_2$ in the middle and $C_3$ on the top.
Within a leaf cluster, vectors are stored as anisotropic product-quantization codes~\cite{avq_2020} with 16 centers per two-dimensional subspace (AH2-LUT16).
The best candidates retrieved from the leaves are rescored against higher-precision vectors.
In high dimensions, we configure \textsc{ScaNN} to route and quantize in a PCA subspace while reranking against full-dimensional vectors.

We also enable SOAR~\cite{soar_2023}, an optimization that stores each vector in a second leaf, chosen so that the two assignments err in different directions, which delivers recall improvements.
We use the default settings for SOAR ($\lambda = 1.5$), except on \textsc{GloVe} and \textsc{Wiki-Cohere} ($\lambda = 1.0$), and on \textsc{DataComp-DR} and \textsc{DBpedia} ($\lambda = 2.0$).
For anisotropic vector quantization, we also use the default weight $\text{AVQ} = 2.5$.
Because the default noise-shaping threshold $\text{NST} = 0.2$ assumes unit-norm vectors, we scale it proportionally to the average vector norm on non-unit-norm datasets ($24.0$ on \textsc{MS-SPACEV}, $100.0$ on \textsc{BigANN}, $2.75$ on \textsc{Wiki-Cohere}, and $2.0$ on \textsc{MS-MARCO}).
We also tune it to $\text{NST} = 0.1$ on the unit-norm datasets \textsc{OpenAI-arXiv}, \textsc{DataComp-DR}, and \textsc{DBpedia}.

Because \textsc{ScaNN} builds its hierarchy bottom-up, we first size the leaf clusters $C_1$ through target memory budgets of 2-bit AH codes: roughly $10\,\text{KB}$ per leaf on billion-scale datasets, $20$ to $64\,\text{KB}$ in high-dimensional PCA space, and $\approx \sqrt{n}$ on small flat indexes.
We train this leaf $k$-means once and build the upper routing levels on the fixed leaf centroids without retraining.
To balance centroid routing overhead against leaf scanning, we scale the routing depth with dataset size.
Small datasets use flat routing, as scanning all leaf centroids directly is fast.
Medium datasets use a two-level tree with $C_2 \approx \sqrt{C_1}$ top-level clusters for a balanced branching factor.
Billion-scale datasets use a three-level tree with a uniform geometric branching factor $B \approx C_1^{1/3}$.
Over these fixed leaf partitions, we evaluate different upper-level routing layouts to select the best configuration without reclustering the dataset.

For reranking, we choose the most compact format that preserves accuracy: \textsf{int8} suffices on \textsc{DBpedia-1536}, \textsc{OpenAI-arXiv}, and \textsc{DBpedia-3072}, and is lossless on \textsc{MS-SPACEV}.
We rerank with \textsf{bf16} on \textsc{BigANN} (where signed \textsf{int8} caps recall at $88.5\%$), on the $d = 768$ inner-product datasets (where \textsf{int8} caps recall at $96\%$), and on the billion-scale float datasets to halve bandwidth with minimal accuracy loss.

At query time, we sweep the number of probed leaves, the number of upper-level clusters visited, and the number of reranked candidates to pick the Pareto-optimal configurations.
To choose the upper-level clusters visited, we sweep over them directly on two-level trees.
On three-level trees, an independent grid over both upper levels is intractable.
Instead, because query latency is dominated by memory bandwidth, we set the upper-level probes as a function of the leaf probes so that each routing level scans a fixed ratio of the leaf bytes.
We evaluate nine routing profiles across three categories: uniform allocations (spending $0.5\times$, $1\times$, or $2\times$ of leaf bytes at both upper levels), mid-heavy allocations ($0.5\times$--$1\times$ at the root and $1\times$--$2\times$ in the middle), and low-routing allocations ($0.05\times$--$0.1\times$ at the root and $0.25\times$--$0.5\times$ in the middle).

\myparagraph{Faiss}
\textsc{Faiss} composes an inverted file (IVF) index from modular pipeline stages~\cite{douze2024faiss}, listed in Table~\ref{tab:construction_params} in factory notation.
The dataset is partitioned into $n$ clusters using $k$-means.
Within each cluster, vectors are split into $m$ subspaces and stored as 4-bit product-quantization codes in a cache-aligned fast-scan layout~\cite{andre2016cache} (\textsf{PQ}$m$\textsf{x4fs}).
To route queries across large numbers of clusters efficiently, \textsc{Faiss} builds an HNSW graph over the $n$ centroids (\textsf{IVF}$n$\textsf{\_HNSW32}).

To balance routing cost against cluster scanning, we sweep the cluster count $n$ around recommended baselines: $\{1{,}024$, $2{,}048$, $4{,}096\}$ on small datasets, $\{16{,}384$, $32{,}768$, $65{,}536$, $98{,}304$, $131{,}072$, $196{,}608$, $262{,}144\}$ on medium datasets, and $\{524{,}288$, $1{,}048{,}576$, $2{,}097{,}152\}$ on billion-scale datasets.
Small datasets use flat routing, as scanning all cluster centroids directly is fast.
Larger datasets route through an \textsf{HNSW32} graph over the centroids.
To keep $k$-means training tractable for tens of thousands to millions of centroids, we use top-down two-level clustering on datasets with $\ge 10\text{M}$ vectors.

An optional OPQ rotation~\cite{ge2013optimized} balances the variance across PQ subspaces and can also project vectors to fewer dimensions.
We evaluate OPQ across different target dimensions, but omit it on \textsc{GloVe} where rotation does not improve accuracy.
In low dimensions, we evaluate two- and four-dimensional OPQ subspaces, where two dimensions per subspace with centroid residuals (\textsf{PQ}$m$\textsf{x4fsr}) performs best.
In high dimensions, we sweep OPQ target dimensions in $\{128, 192, 256, 384, 512\}$, and store uncentered vectors because centroid subtraction degrades throughput--recall performance.

The best candidates retrieved from the clusters are rescored against higher-precision vectors.
We choose the most compact format that preserves accuracy: 8-bit scalar quantization (\textsf{SQ8}) suffices on \textsc{DBpedia-1536}, \textsc{OpenAI-arXiv}, and \textsc{DBpedia-3072}, and is lossless on the two integer datasets.
We rerank with \textsf{bfloat16} (\textsf{SQbf16}) on the $d = 768$ inner-product datasets and the billion-scale float datasets.
On \textsc{GloVe}, we rerank against exact floats (\textsf{RFlat}).

At query time, we sweep the number of probed clusters and the number of reranked candidates to pick the Pareto-optimal configurations.
When using an HNSW coarse quantizer, we additionally sweep the graph search depth (\textsf{efSearch}) to balance centroid routing latency against cluster retrieval recall.

\clearpage

\section{Library Versions}
\label{sec:library-versions}

\begin{table}[h]
  \centering
  \footnotesize
  \caption{Evaluated libraries, versions, and Git commit hashes.}
  \label{tab:library_versions}
  \begin{tabular}{lll}
    \toprule
    Library & Version & Git Commit \\
    \midrule
    \textsc{Faiss}~\cite{douze2024faiss} & \textsf{v1.15.0} & \href{https://github.com/facebookresearch/faiss/tree/20f14b31a6d54e243a3d1de6ae193fc4c3ec18ed}{\texttt{20f14b}} \\
    \textsc{Hnswlib}~\cite{hnswlib} & \textsf{v0.9.0} & \href{https://github.com/nmslib/hnswlib/tree/d9b3608c83d83b46c96e25088cb1d729b29dcfe9}{\texttt{d9b360}} \\
    \textsc{ParlayANN}~\cite{manohar2024parlayann,parlayann_repo} & -- & \href{https://github.com/cmuparlay/ParlayANN/tree/a48c6a34e5f959190992aa531cc18454cfcc3dd3}{\texttt{a48c6a}} \\
    \textsc{SVS}~\cite{aguerrebere2023similarity} & \textsf{v0.4.0} & \href{https://github.com/intel/ScalableVectorSearch/tree/75e30f2}{\texttt{75e30f}} \\
    \textsc{DiskANN}~\cite{krishnaswamy2024diskann} & \textsf{v0.59.0} & \href{https://github.com/microsoft/DiskANN/tree/c84530e57c0305a2502c35770301ce5667f265f7}{\texttt{c84530}} \\
    \textsc{ScaNN}~\cite{avq_2020,soar_2023} & \textsf{v1.4.3} & \href{https://github.com/google-research/google-research/tree/807d4a2f41202059bac2446259d135a89ed3630a/scann}{\texttt{807d4a}} \\
    \textsc{SymphonyQG}~\cite{gou2025symphonyqg,gao2025rabitq} & \textsf{v0.3.4} & \href{https://github.com/VectorDB-NTU/RaBitQ-Library/tree/1182cbdf3b05fc6c00fb2061dd0f396822b66c9d}{\texttt{1182cb}} \\
    \bottomrule
  \end{tabular}
\end{table}

\end{document}